# Thermodynamic evolution of a sigmoidal active region with associated flares

Sargam M. Mulay[1]★ , Durgesh Tripathi[2] , and Helen Mason[3]

[1]*School of Physics & Astronomy, University of Glasgow, G12 8QQ, Glasgow, UK*
[2]*Inter-University Centre for Astronomy and Astrophysics, Ganeshkhind, Pune 411007, India.*
[3]*DAMTP, Centre for Mathematical Sciences, University of Cambridge, Wilberforce Road, Cambridge, CB3 0WA, UK*

## ABSTRACT

Active regions often show S-shaped structures in the corona called sigmoids. These are highly sheared and twisted loops formed along the polarity inversion line. They are considered to be one of the best pre-eruption signatures for CMEs. Here, we investigate the thermodynamic evolution of an on-disk sigmoid observed during December 24-28, 2015. For this purpose, we have employed Emission Measure (EM) and filter-ratio techniques on the observations recorded by the Atmospheric Imaging Assembly (AIA) onboard the Solar Dynamics Observatory (SDO) and X-ray Telescope (XRT) onboard Hinode. The EM analysis showed multi-thermal plasma along the sigmoid and provided a peak temperature of ∼10-12.5 MK for all observed flares. The sigmoidal structure showed emission from Fe xviii (93.93 Å) and Fe xxi (128.75 Å) lines in the AIA 94 and 131 Å channels, respectively. Our results show that the hot plasma is often confined to very hot strands. The temperature obtained from the EM analysis was found to be in good agreement with that obtained using the XRT, AIA and GOES filter-ratio methods. These results provide important constraints for the thermodynamic modelling of sigmoidal structures in the core of active regions. Moreover, this study also benchmarks different techniques available for temperature estimation in solar coronal structures.

## 1 INTRODUCTION

On disk observations of the Sun's corona in extreme ultraviolet (EUV) and X-rays often exhibit *S-shaped* (reverse S-shaped and/or two J-shaped) loops embedded in the core of active regions. These sheared and twisted coronal structures are known as *Sigmoids* (Rust & Kumar 1996; Gibson et al. 2002; Schwenn et al. 2006; Tripathi et al. 2006b, 2009) and could be considered as manifestations of twisted flux tubes (Gibson & Fan 2006; Chen 2017; Inoue et al. 2018). They are often seen on-disk in the lower corona prior to large-scale plasma eruptions such as Coronal Mass Ejections (CMEs) (Canfield et al. 1999; Green et al. 2002; Tripathi et al. 2004; Webb et al. 2013; Green & Kliem 2014). It has been suggested that a twist in the flux tubes could be introduced by a shearing motion of the footpoints of flux tubes (Gibson et al. 2004; Green, L. M. et al. 2011), which destabilizes the sigmoidal structure and eventually leads to an eruption (often accompanied by solar flares and CMEs). The dynamic events, such as sigmoid eruptions along with CMEs, are capable of transferring mass and energy, including energetic particles to the heliosphere, and play an important role in shaping the space weather and interplanetary environment (Gibson et al. 2006; Kilpua et al. 2017; Palmerio et al. 2018).

Most of the observations show that flux ropes carry cold dense filaments/prominences (Chifor et al. 2006, 2007; Green et al. 2007) together with hot sigmoids (Gibson et al. 2002; Del Zanna et al. 2002; Tripathi et al. 2006b) with them, which appear as a bright core during the mass ejection. Therefore, sigmoids are key elements of CME eruptions and could be used to predict the onset of a CME. Observationally, sigmoids have been considered as tracers for flux ropes (Canfield et al. 1999; James et al. 2017).

A number of studies are dedicated to an understanding of the formation and evolution of sigmoids as well as associated flux rope structures. Fan (2010) carried out a three-dimensional MagnetoHydrodynamic (MHD) simulations of the evolution of the magnetic field in the corona. The authors observed the existence of a sigmoid-shaped current layer underlying the flux rope during the pre-eruption quasi-static stage. The magnetic energy dissipation in the current layer leads to the heating of the field lines. Observationally, this resembles the heating of the quiescent X-ray sigmoid loops which are generally seen prior to a CME eruption.

Jiang et al. (2013) studied MHD simulations of the initiation of solar eruptions by taking the initial conditions from a nonlinear force-free field (NLFFF) extrapolation prior to a sigmoid eruption. The study revealed that the eruption occurrence was due to the initiation of a torus instability at the flux rope, which is a result of reconnection at the null point present in the core of the sigmoid. Further comprehensive studies by Jiang et al. (2014) constrained the NLFFF using vector magnetograms and reproduced an evolving sigmoidal flux rope which perfectly matched with observations. A data-constrained high-resolution MHD simulation of a sigmoidal active region by Jiang et al. (2018) provided a realistic view of the magnetic configuration and three-stage magnetic reconnection scenario involved in a CME initiation process.

A number of studies involve an understanding of the physical parameters (such as electron number density, temperature, Doppler velocities, etc.) of sigmoids. Based on spectroscopic observations taken with the Coronal Diagnostic Spectrometer (CDS) (Harrison et al. 1995) onboard Solar and Heliospheric Observatory (SoHO) (Domingo et al. 1995), Tripathi et al. (2006b) reported a temperature ∼2 MK (using Fe xvi and Si xii lines) and an electron density of $9.1 \times 10^8$ cm$^{-3}$. In contrast, Del Zanna et al. (2002) studied a flare associated with a sigmoidal brightening and reported an electron number density of $2.5$-$7 \times 10^{11}$ cm$^{-3}$ obtained using a transition region line from O iv, a high-temperature of 8 MK using the Fe xix line and strong blue-shifts of ∼30 km/s (using Mg x and Si xii lines). The spectroscopic study by Gibson et al. (2002) reported the presence of multithermal plasma with a temperature $1$-$3 \times 10^5$ K and coronal plasma between 2 and 7 MK along the sigmoid. During the active phase of the sigmoid, the temperature rose to 8 MK. From these observations, they confirmed that the hotter plasma lies above the cooler plasma and found that it is consistent with the magnetic field extrapolation. Cheng et al. (2014) and James et al. (2018) reported higher temperatures ranging between 7–14 MK using a differential emission measure (DEM) analysis of EUV images.

Multiple transient brightenings have been observed along the sigmoidal structure (Gibson et al. 2002; Sterling et al. 2000). The S-shape of the soft X-ray sigmoid was only revealed in EUV images during these brightenings. Therefore, it is important to study these brightenings which enable the sigmoid to sustain a high temperature and stand out from the ambient plasma. The question arises whether these transient brightening events are associated with small X-ray flares and how hot the sigmoid can get during these events. We further note that there are number of studies that emphasize the roles of small scale transient brightenings that leads to the formation and heating of sigmoidal structures before its eventual eruption (Tripathi et al. 2009; James et al. 2020; Kliem et al. 2021).

It is important to note that although imaging and spectroscopic observations of individual sigmoidal regions have been studied previously, the thermal distribution of plasma along a sigmoidal region and its evolution has not yet been studied fully. This is primarily due to the lack of temperature coverage in the observations. By taking advantage of high spatial and temporal resolution observations from the Atmospheric Imaging Assembly (AIA; Lemen et al. 2012) and X-ray Telescope (XRT; Golub et al. 2007; Kano et al. 2008) instruments, we have focused on the thermal structure of an on-disk active region sigmoid and studied its heating/cooling during different phases of solar flares. The main aim of this paper is to study flare-related sigmoids in EUV and X-ray images using filter-ratio and Emission Measure methods. The purpose is to study how hot sigmoids can get during different classes of small X-ray flares. We also study how well different techniques perform for temperature estimation perform.

Initially, we have carried out a survey of sigmoidal regions observed by XRT during 2010-2019. By making use of XRT full disk images available on the Solar Monitor website[1] and the Heliophysics Events Knowledgebase[2] (HEK; Hurlburt et al. 2012), we have identified sigmoidal active regions and created a list. We chose the sigmoid observed in the NOAA active region #12473 for the detailed study presented here, as it was located near the disk centre and produced a number of flares during the course of evolution. According to the sigmoid classification by Gibson et al. (2006), this is a 'persistent' sigmoid structure which consists of a bundle of sheared S-shaped loops and this S-shape lasted for five days before the eruption.

This paper is structured as follows. We present the UV/EUV and X-ray imaging observations in §2. The thermodynamic structure of the sigmoid and physical parameters are determined in §3. Finally, we summarise and discuss our results in §4.

## 2  OBSERVATION AND DATA

We study a sigmoid structure associated with the NOAA active region #12473 during its transit across the central meridian from Dec. 24 to 28, 2015. For this study, we have used the observations recorded from the Geostationary Environmental Operational Satellite (GOES–15), AIA and Helioseismic and Magnetic Imager (HMI; Scherrer et al. 2012; Schou et al. 2012) on board Solar Dynamics Observatory (SDO; Pesnell et al. 2012) and XRT on board the Hinode (Kosugi et al. 2007) satellite. GOES provides full Sun integrated X-ray fluxes that help us identify the flares. AIA images the Sun in UV and EUV using nine filters with a pixel size of 0.6″ and cadence of 12 s, whereas HMI provides the measurement of magnetic fields on the surface of the Sun with a pixel size of 0.5″ and a time cadence of 45 s. In this study, we have only used the line-of-sight (LOS) magnetic field. The XRT instrument provides X-ray images of the Sun in different filters with a pixel size of 1.02″ and temporal resolution of ∼1-3 min. To study the chromospheric activity, in addition to 1600 Å images from AIA, we have examined Ca ii H 3969 Å images from the Broadband Filter Imager (BFI) (with pixel size of 0.054″ and time resolution of 3 min) board the Solar Optical Telescope (SOT; Suematsu et al. 2008; Tsuneta et al. 2008) onboard Hinode.

During this period, GOES observed 16 flares, viz. 4 B-class, 10 C-class, and 2 M-class. We have studied all the C-class and M-class flares and the thermodynamic evolution of the sigmoid during these events. However, we have discarded the B-class flares due to the low signal-to-noise-ratio (SNR). The X-ray fluxes for all 16 flares along with their temperatures obtained using GOES are shown in Fig. 1. Here, we discuss the evolution of the sigmoid during the C1.6 flare, which was observed on Dec. 26, 2015, between 01:30 and 3:00 UT and the M1.8 flare observed on Dec. 28, 2015, between 11:00 and 13:30 UT associated with the final eruption of the sigmoid. The details of all C and M-class flares are given in Table A. The two flares discussed below in detail are marked as #8 and #12 in the table. The relevant results for other flares are presented in §3.5 and Appendix A.

We have downloaded full disk AIA level 1 data and converted this to level 1.5 using the `aia_prep.pro` from the SolarSoftWare (SSW; Freeland & Handy 1998) library. All the images were mapped to a common plate scale and co-aligned. The intensities were normalised with their respective exposure times. Since this study is focused on the thermodynamics of the dynamic events, it is important to note that the intensities measured by AIA channels are multi-thermal (O'Dwyer et al. 2010, Boerner et al. 2012 Del Zanna et al. 2011a, 2013) 94 Å (Fe x, Fe xiv, Fe xviii), 131 Å (Fe viii, Fe xxi), 171 Å (Fe ix), 193 Å (Fe xii, Ca xvii, Fe xxiv), 211 Å (Fe xiv), 304 Å (He ii), 335 Å (Fe xvi), 1600 Å (C iv+continuum) and 1700 Å (dominated by C i multiplet at 1656 Å and He ii 1640 Å with negligible contribution from the continuum). The dominant contributions from ions in each of the AIA channels are given in parentheses. Based on a thorough investigation using EUV imaging and spectroscopic data, Del Zanna (2013) and Warren et al. (2012) provided empirical formulae to estimate the main contribution of Fe xviii 93.93 Å line in the AIA 94 Å channel. Both methods provide similar contribution of Fe xviii emission. In this paper, we have used the empirical formula by Del Zanna (2013) (see Eq. 1). The AIA 171 and 211 Å channels are used to remove the low-temperature component from the AIA 94 Å channel.

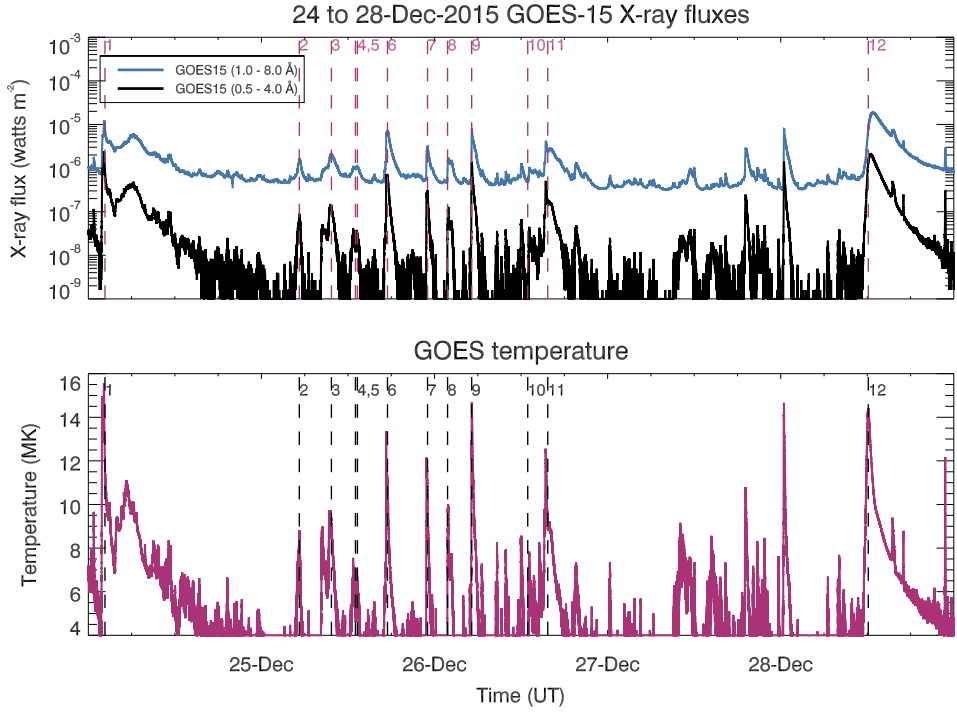

**Figure 1.** Top panel: GOES X-ray fluxes in 1.0-0.8 Å (blue) and 0.5-4.0 Å (black) channels. Bottom panel: the temperatures obtained from two GOES channels for the duration of Dec. 24-28, 2015. The dashed lines indicate the peak timings of flares studied. The numbers indicate the flare events same as given in Table A.

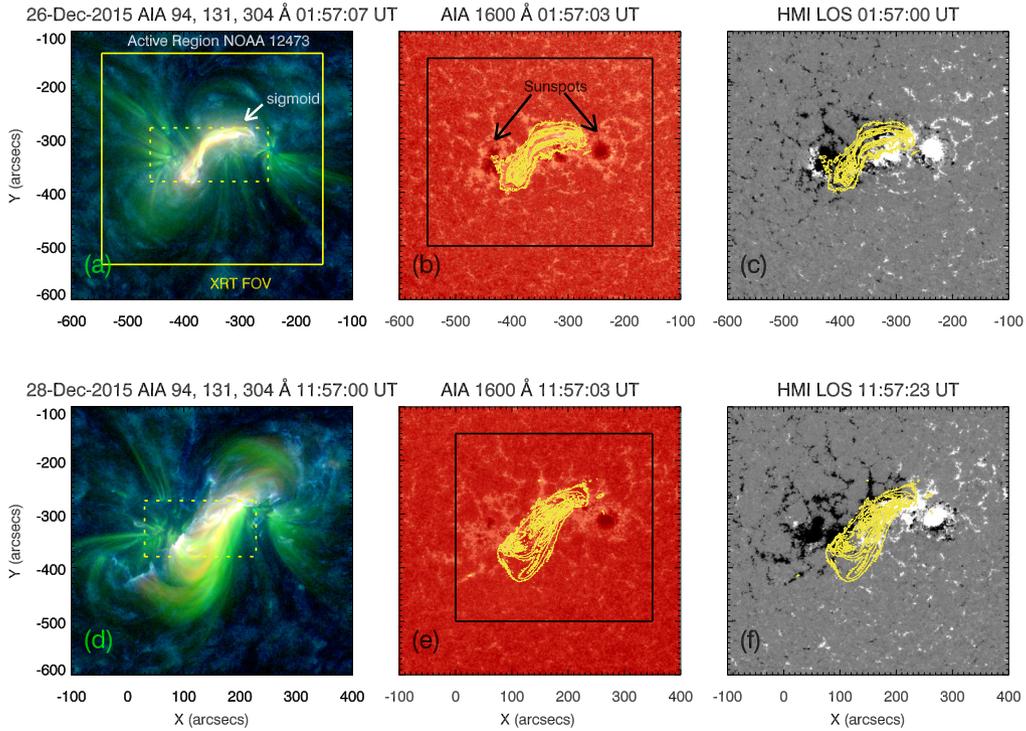

**Figure 2.** The AIA images of the sigmoid during the initial (top row) and the eruption phases (bottom row). Panels (a) and (d): composite images of an active region NOAA #12473 created using the AIA 94 (red), 131 (green) and 304 Å (blue) channels. The solid (dashed) yellow boxed region represents the XRT (SOT) FOV. Panels (b) and (e): AIA 1600 Å images. The two associated sunspots are marked by black arrows. The over-plotted black boxed regions were used to create AIA light curves. Panels (c) and (f): HMI LOS magnetograms. The over-plotted contours of the sigmoid and it's eruption phase are obtained from AIA 94 Å images.

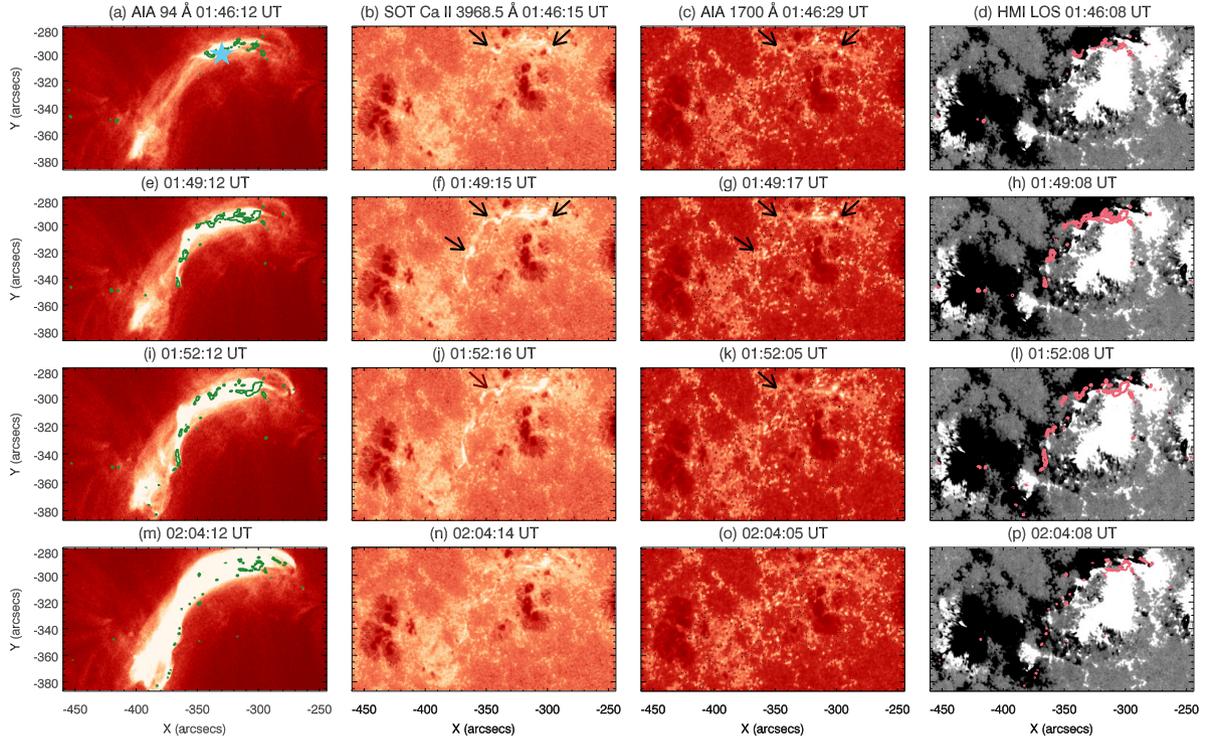

**Figure 3.** The evolution of sigmoid structure on Dec. 26, 2015 observed by various instruments as labelled. The FOV shown corresponds to the yellow dashed boxed marked in the panel (a) of Fig. 2. First column: AIA 94 Å images of the sigmoid. The blue star indicates the location (at X = -330″ and Y = -300″) where the first brightening occurred. Second and third columns: the chromospheric counterpart of the sigmoid as observed by SOT in Ca II 3968.5 Å and by AIA in 1700 Å images respectively. The brightenings observed in SOT images are shown by black arrows in AIA 1700 Å images. Fourth column: evolution of the photospheric LOS magnetic flux scaled within ±100 G as observed by the HMI. The location of brightenings along sigmoid in Ca II images are shown as green and pink contours in the first and fourth column images respectively.

$$I(\text{Fe XVIII}) = I(AIA\ 94) - I(AIA\ 211)/120 - I(AIA\ 171)/450 \quad (1)$$

Figure 2 displays the sigmoid structure (panel (a)) (see also Movie 1, which is available as supplementary material) and its eruption (panel (d)) (see also Movie 2, which is available as supplementary material). Panels (a) and (d) are composites of the images recorded in AIA 94 (red), 131 (green) and 304 Å (blue) channels. The corresponding 1600 Å images are shown in panels (b) and (e). Overplotted are the intensity contours obtained from the corresponding AIA 94 Å image. The over-plotted yellow box with solid line (dashed lines) in panels (a) and (d) are the XRT (SOT) field-of-view (FOV). The over-plotted boxes in panels (b) and (e) with black solid lines represent the region that was chosen for the study of the averaged light curves. The sigmoidal region and associated sunspots are labelled.

The active region consisted of a positive polarity leading sunspot and negative polarity trailing sunspot with a Hale class of $\beta\gamma$ magnetic configuration[3] (also see images in column 4 of Figs. 3 and 4 i.e. in panels (d), (h), (i), and (p)). The co-alignment was carried out between the AIA 1700 Å images and the LOS magnetograms by taking the sunspots as a reference. A comparison between the HMI and AIA images shows that the sigmoid was located along the polarity inversion line (PIL) between the two sunspots.

[3] See https://www.solarmonitor.org/?date=20151226.

The XRT observed the sigmoid using Al-poly and Be-thin filters. We have studied the XRT observations on Dec. 26, 2015, between 1:30 and 3:00 UT, though there was a data gap between 02:09 and 02:34 UT. However, there were no XRT data available during the sigmoid eruption on Dec. 28, 2015. For our analysis, we have used the XRT calibration by Narukage et al. (2011). We have co-aligned the XRT images with AIA images taken at 335 Å using the procedure of Yoshimura & McKenzie (2015).

## 3 MULTI-WAVELENGTH EVOLUTION OF THE SIGMOID

On Dec. 26, 2015 (corresponding to the first row in Fig. 2), we examined the evolution of the sigmoid at multiple wavelengths (see also Movie 1, which is available as supplementary material). Fig. 3, first column displays AIA 94 Å images taken at various instances. The Ca II and 1600 Å images are shown in the second and third columns respectively. The evolution of the LOS magnetic flux density is displayed in the fourth column.

Monitoring the EUV images taken by AIA in 94 and 131 Å, we observed a slight movement of the plasma at around 01:45:12 UT at [X = -330″, Y = -300″] marked with a blue star in panel (a). With passing time, the brightening started to spread symmetrically in both directions. Until 01:49:12 UT, the brightening was confined to a small area of the sigmoid structure (see panel (e)). Soon after, it started to spread over the inner edge/boundary of the sigmoid i.e., the side of the sigmoid, which is close to the leading sunspot, until 01:52:12 UT (see panel (i)). The entire area of the sigmoid had brightened by 02:04:12 UT (panel (m)). No other activity was observed during this period. From 02:36 UT, we observed a decrease in the intensity along the sigmoidal structure and no bright structure was visible by 03:00 UT. The whole process of brightening and fading, possibly corresponding to heating and cooling took place within 75 minutes. A similar phenomena was also observed in the AIA 131 Å channel.

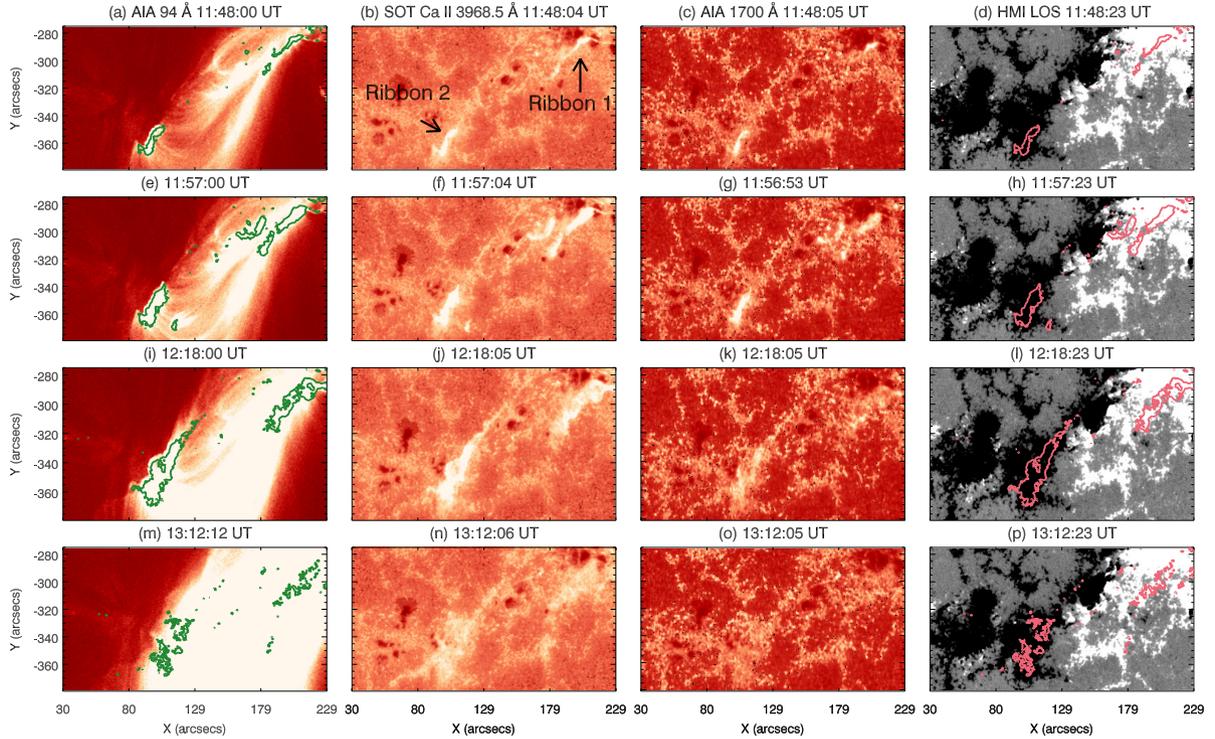

**Figure 4.** The sigmoid observed during the eruption phase on Dec. 28, 2015. The FOV shown in these images is the same as those shown as a yellow dashed boxed region in the panel (d) of Fig. 2. Same as Fig. 3, the AIA 94 Å, SOT and AIA 1700 Å and HMI images are shown in Columns 1-4.

We examined the chromospheric activity using Ca II 3969 Å and AIA 1700 Å images (see columns 2 and 3 - panels (b), (f), (j) and (n); (c), (g), (k) and (o) of Fig. 3). A similar brightening at the same location marked in panel (a) was observed in Ca II, AIA 1600 and 1700 Å. The first image of the brightening was available at 01:46:15 UT in Ca II. Since successive images recorded in Ca II are ~3 min apart, it is difficult to determine whether the brightening occurred in chromosphere heated at 01:46 UT or earlier. However, the high cadence AIA 1600 and 1700 Å images confirmed that the brightening started at 01:45:51 UT. In a similar way to the coronal activity, the brightening moved in the South-West direction as well as in the horizontal direction towards the north of the leading sunspot. The brightening is shown with black arrows in AIA 1700 Å and Ca II images and the contours of this brightening are over plotted on AIA 94 Å and LOS HMI magnetograms. Finally, the brightening disappeared after 02:04 UT.

On Dec. 28, 2015 (corresponding to the bottom row of Fig. 2), the sigmoidal region approached the disk center and showed elongated S-shaped loops in the core of the active region (see also Movie 2, which is available as supplementary material). The upper ends of the loops were anchored in a positive polarity (north of the positive polarity sunspot) whereas the lower ends were anchored in a negative polarity region (south of the negative polarity sunspot) (see panel (f) of Fig. 2). We call these regions (endpoints of the loops) 'footpoints'. The bundle of twisted loops was distinctly visible in the AIA 94 Å channel and images are shown in column 1 (panels (a), (e), (i) and (m)) of Fig. 4. The first brightening was observed at 11:32:36 UT at a similar location to that where we observed the first brightening along the sigmoid on Dec. 26, 2015. We confirmed the location by de-rotating the AIA 94 Å images. The brightening slowly spread over a small area, which was a part of the sigmoidal structure and located at the north end. Thereafter, we observed that there is a formation of new loops followed by the brightening at 11:48 UT at the footpoints of the loops (see panels (e) and (i) of Fig. 4). These brightenings were seen to be moving away from each other, indicating the movement of the footpoints of loops. The loops along the sigmoidal structure were highly sheared and that probably led to the sigmoid eruption. A new bundle of elongated post eruption arcades (Tripathi et al. 2004) started to form between two sunspots. These had a similar morphology to the sigmoid. This is nicely seen in the AIA 171 and 211 Å channel images. A new set of loops along the sigmoid started to grow and diffuse emission started to fill the region with intense brightening at the loop tops (see also Movie 2, which is available as supplementary material). A similar phenomenon was observed in the AIA 131 Å channel.

The images shown in the columns 2 & 3 of Fig. 4, displays Ca II and AIA 1700 Å images, respectively. As we can see, during this flare (unlike the C1.6 flare), the chromospheric brightenings were observed at the footpoints of the sigmoidal loops. The two bright ribbons, one close to the leading sunspot and other to the trailing one, were observed at the footpoints of the sheared loops forming the sigmoidal structure, similar to the observations presented by Tripathi et al. (2009). We call these ribbons 'Ribbon 1' and 'Ribbon 2', respectively.

The first chromospheric brightening was observed in Ca II images at 11:30 UT at the ribbon 2. The ribbon 1 started to brighten up about 10 minutes later between 11:39 and 11:42 UT. Initially, ribbon 2 was brighter than ribbon 1, but both ribbons appeared similarly bright at 11:54 UT. A small loop like brightening was observed along the PIL, which is clearly seen in the Ca II image at 11:57 UT (see panel (f)) and AIA 1700 Å image (see panel (g)). The contours of these brightenings are plotted on the AIA 94 Å images as well as on the HMI LOS magnetogram images. A similar brightening at the loop along the PIL was also seen in AIA 94 Å image at 11:57:00 UT (see panel (e)). The brightening fades away by 12:03 UT. With time the brightenings along the two ribbons started to move away from each other (away from the PIL) and spread over the nearby region. Similar activity was also observed in the hotter AIA 131 Å channels, which showed a signature of the sigmoid eruption. This activity started at 11:51 UT and lasted until 12:45 UT. The ribbon 1 started to become less bright as compared to ribbon 2 as it moved away but the brightening along the ribbon 1 started to grow horizontally as well as towards north.

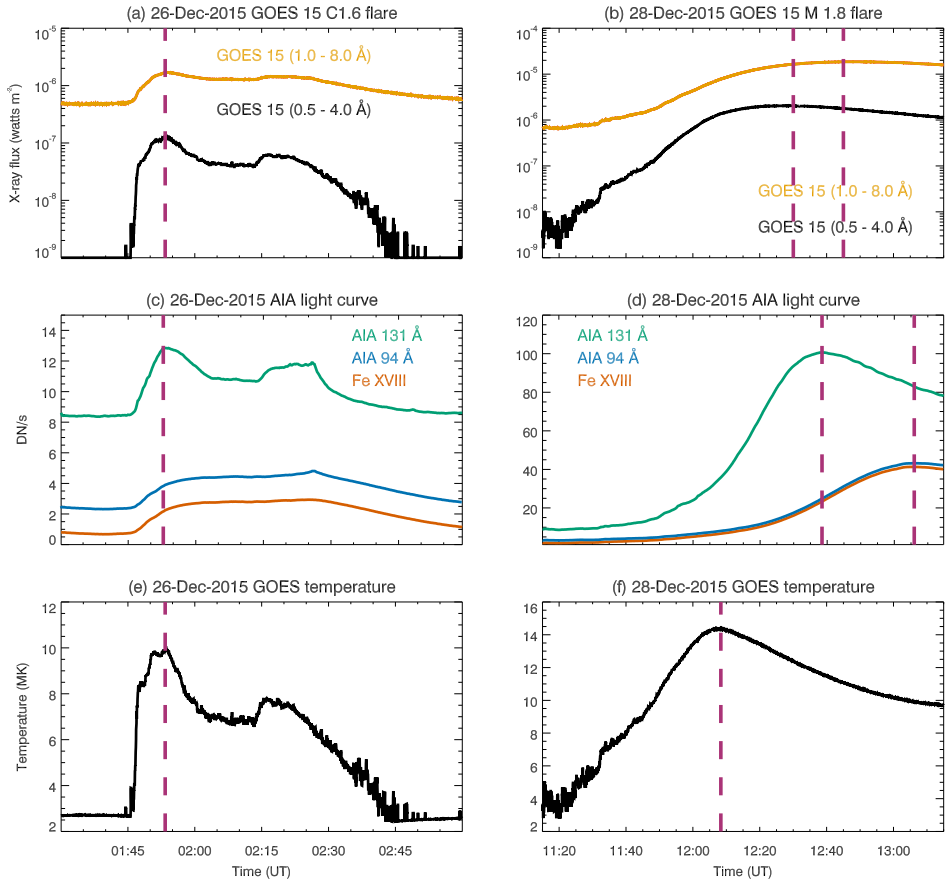

**Figure 5.** Top panels: light curves of X-ray fluxes during the C1.6 (panel (a)) and M1.8 (panel (b)) flares observed by GOES 15 in 1.0-0.8 Å and 0.5-4.0 Å channels. Middle panels: the AIA light curves averaged over the regions marked by black boxed regions in panels (b) and (e) of Fig. 2. Bottom panels: GOES temperature profiles for C1.6 and M1.8 X-ray flares. The dashed vertical lines indicate the timings of the peak for each of the curves.

In order to study the activity along the sigmoid structure on both the days, we inspected X-ray fluxes to check whether the EUV brightening had any connection with activity recorded by GOES. For this purpose, we obtained the X-ray flux light curves using 1.0-8.0 Å and 0.5-4.0 Å bands on GOES-15 (see panels (a) and (b) in Fig. 5). According to the National Oceanic and Atmospheric Administration's (NOAA) Space Weather Prediction Center (SWPC), the X-ray fluxes (in GOES 1.0–8.0 Å channel) show that the C1.6 flare that occurred on Dec. 26, 2015 started at 01:44 UT, peaked at 01:53 UT and ended at 02:32 UT. Similarly, the M1.8 flare that occurred on

Dec. 28, 2015 started at 11:20 UT, peaked at 12:45 UT and ended at 13:09 UT. The rising phase of a flare is defined as the time from the start to the peak of the X-ray flux in GOES 1.0-8.0 Å channel. It is observed that the rise time for the C1.6 flare (~9 min) was smaller in comparison to that for the M1.8 flare (1 hour 25 min). Interestingly, during the M1.8 flare, the GOES 0.5-4.0 Å channel showed a peak (at 12:30 UT) almost 15 min earlier than the peak in the 1.0-8.0 Å channel (at 12:45 UT). There was no time difference recorded in the peak timings of X-ray channels for C1.6 flare.

Since GOES provides full disk-averaged X-ray fluxes, the spatial information of the flare location cannot be determined directly by using these flux profiles. To investigate this, we obtained the light curves for AIA 131 and 94 Å channels and Fe XVIII images over the box regions shown in panels (b) and (e) of Fig. 2. These light curves for both C-class and M-class flares are shown in panels (c) and (d) of Fig. 5. The plots reveal that the count rates (DN/s) in the AIA 94 Å channel were smaller compared to those recorded by the AIA 131 Å channel for both the flares. We have also isolated Fe XVIII emission using the procedure described earlier and plotted the light curves in panels (c) and (d). We note that a significant amount of Fe XVIII emission was present at the sigmoidal location.

Further, we compared the AIA light curves with those of GOES. The nature of the curves obtained from AIA was very similar to those we obtained from the GOES channels. Therefore, we confirmed that both X-ray flares were associated with sigmoidal activity.

In the case of the C1.6 flare, the AIA 131 Å channel showed a peak at 01:52:30 UT i.e. ~30 sec earlier than the GOES peak (at 01:53 UT in 1-8 Å channel), whereas the emission in the AIA 94 Å channel as well as Fe XVIII showed a gradual increase in the intensity. Similarly, in the case of M1.8 flare, the AIA 131 Å channel showed a peak at 12:38 UT i.e. ~7 min earlier than the GOES peak (at 12:45 UT in 1-8 Å channel) whereas the AIA 94 Å showed a peak much later at 13:06 UT (later than GOES flare peak). The systematic peaks indicate the heating in AIA 131 Å channel (mostly emission from Fe XXI at temperature 10 MK) and further cooling of plasma to a temperature of ~8 MK as seen in AIA 94 Å channel (mostly emission from Fe XVIII) (see section 3.2 for further details).

By comparing light curves for the derived Fe XVIII emission and AIA 94 Å, we found that some cool component (weak contributions from Mg VIII and Fe VIII lines) was likely to have been present at the flaring location during all phases of the C1.6 flare whereas in the case of the M1.8 flare, AIA 94 Å was mostly dominated by Fe XVIII emission and the cool component was only present initially during the impulsive phase of the flare. This is also confirmed using the EM analysis described in section 3.2.

Keeping in mind that the prime aim of this work is to study the thermodynamic evolution of sigmoidal regions, we have measured temperatures in a sigmoid at various instants of time using different techniques. The study was carried out for the purpose of benchmarking different methods.

There are essentially two ways of obtaining the temperature of the plasma, namely an Emission Measure (EM) analysis and filter ratio methods. The former works under the assumption that there is multi-thermal plasma along the LOS, whereas the later works on the assumption of iso-thermal plasma along the LOS. We have applied filter ratio techniques on both AIA as well as XRT observations and an EM method to AIA observations. Considering the known constraints on both the methods, we compared the results obtained from these methods and made an attempt to get a reliable distribution of temperatures of the flaring sigmoidal region.

As an additional constraint, we have obtained temperatures from the two GOES X-ray channels and have compared the results with those obtained from the EM and filter-ratio analyses. Since the temperatures obtained using the GOES observations is based on the full disk integrated X-ray flux, we study these first and then in later sections we move towards the EM and filter ratios.

### 3.1 Temperature Estimates using GOES X-ray Fluxes

Using the observed ratio of responses in the two GOES channels namely 0.5–4.0 Å and 1.0–8.0 Å, one can obtain the temperature of an isothermal plasma using the GOES package available in the SSW library (Freeland & Handy 1998). These methods were originally developed by Thomas et al. (1985) and Garcia (1994). For this purpose, we have used the light curves shown in Fig. 5 (panels (a) and (b)) for the two flares. The derived temperature profiles are displayed in panel (e) and (f) of Fig. 5.

For the C1.6 flare observed on Dec. 26, 2015, the initial temperature along the sigmoid during the pre-flare phase was ~2.5 MK. Then the temperature began to increase rapidly during the impulsive phase (01:44 UT onward) and recorded a rise from 2.5 to 8 MK. The GOES temperature profile showed two intermediate peaks at ~8.5 and ~9.5 MK before it reaches a peak temperature of 10 MK at 01:53 UT. The peak temperature was attained in 9 minutes from the start. This is in good agreement with the AIA 131 Å light curve (see panel (c) of Fig. 5). During the decay phase, the plasma started to cool down slowly and finally reached 7 MK at 02:12 UT in 19 minutes. There is another small peak observed in the temperature profile at 02:15 UT at the same time as a rise in emission was seen with the GOES X-ray flux profile. At this time the temperature again increased to ~7.5 MK and slowly cooled down to 2.5 MK by 02:45 UT. This is consistent with the Fe XVIII emission observed in the AIA light curves (see panel (c) of Fig. 5).

For the M-class flare, however, even before the start of the flare, GOES recorded a temperature of about ~4 MK. The flare started to rise at 11:20 UT, reached a peak temperature of ~15 MK in 48 minutes. The GOES recorded a peak temperature at 12:08 UT (see panel (f) of Fig. 5), which is ~38 min (~23 min) earlier than the peak of flare emission recorded in 1.0-8.0 Å (0.5-4.0 Å) channel. We compared the timings and temperature with AIA 131 Å (94) light curve and found that the GOES showed a slightly higher temperature ~12 MK (10 MK) than the peak of Fe XXI (Fe XVIII) line considering that these ions dominate the AIA 131 (94) Å channel (later this is confirmed in section 3.2). The GOES temperature started to decrease rapidly after 12:10 UT and reached 10 MK at 13:00 UT in 50 minutes.

### 3.2 Emission Measure analysis using AIA imaging data

The thermodynamic nature of the sigmoid was studied by performing an EM analysis on the EUV images recorded with AIA in the 94, 131, 171, 193, 211 and 335 Å channels. The EM inversion method developed by Cheung et al. (2015) was used to study the temperature distribution along the sigmoidal structure during various phases of both flares.

For this purpose, we chose the unsaturated (with $DN$<16000 per pixel) AIA EUV images. The temperature responses for the AIA channels were computed using the SSW routine `aia_get_response.pro`. The CHIANTI atomic database (Dere et al. 1997; Landi et al. 2013), ionization equilibrium calculations, the electron number density of $1 \times 10^{10}$ cm$^{-3}$ and coronal abundances (by combining the abundances from Feldman 1992; Grevesse & Sauval 1998; Landi et al. 2002) were used in this analysis. The temperature range of log $T$ [K] = 5.4 - 7.5 (0.2–31 MK) was chosen to compute the EM per temperature bin ($\triangle T$ = 0.1) using normalised intensities per pixel (DN/s/pix) in six EUV channels.

We created EM maps for the sigmoidal active region in various temperature intervals during the C1.6 flare (see also Movie 3, which is available as supplementary material). The EM maps obtained at GOES flare peak timing (01:53 UT) are displayed in Fig. 6. The plots (and associated animation) clearly reveal the presence of multi-thermal plasma distributed in the active region, including along the sigmoidal region. Within the active region, the region outside sigmoid shows plasma at temperature log $T$ [K] < 6.5 (3.1 MK). The emission from the sigmoidal structure, however, is dominated by plasma at log $T$ [K] > 6.5 (3.1 MK) and has minimal contribution from low-temperature (log $T$ [K] < 6.3 (<2 MK)).

We note red patches in the EM maps obtained between log $T$ [K] = 6.5 to 7.3 displayed in panels (c), (d), (e) and (f) of Fig. 6. Such patches are observed during the impulsive phase and at the peak of the flare, and are located at the same region where we observed the first brightening at 01:46 UT in the AIA 94 Å channel. The patch is strongest in the EM maps obtained with log $T$ [K] = 6.7 - 6.9 (5–8 MK). At the highest temperature, only a few strands along the sigmoid were seen at higher temperatures >12.5 MK (see panel (f)). During the decay phase of the flare, EM values were seen to be decreasing rapidly in that region and the plasma started to cool down. This is evident from the lower temperature EM maps where high EM (red patches) are seen at later times. The EM analysis provides a

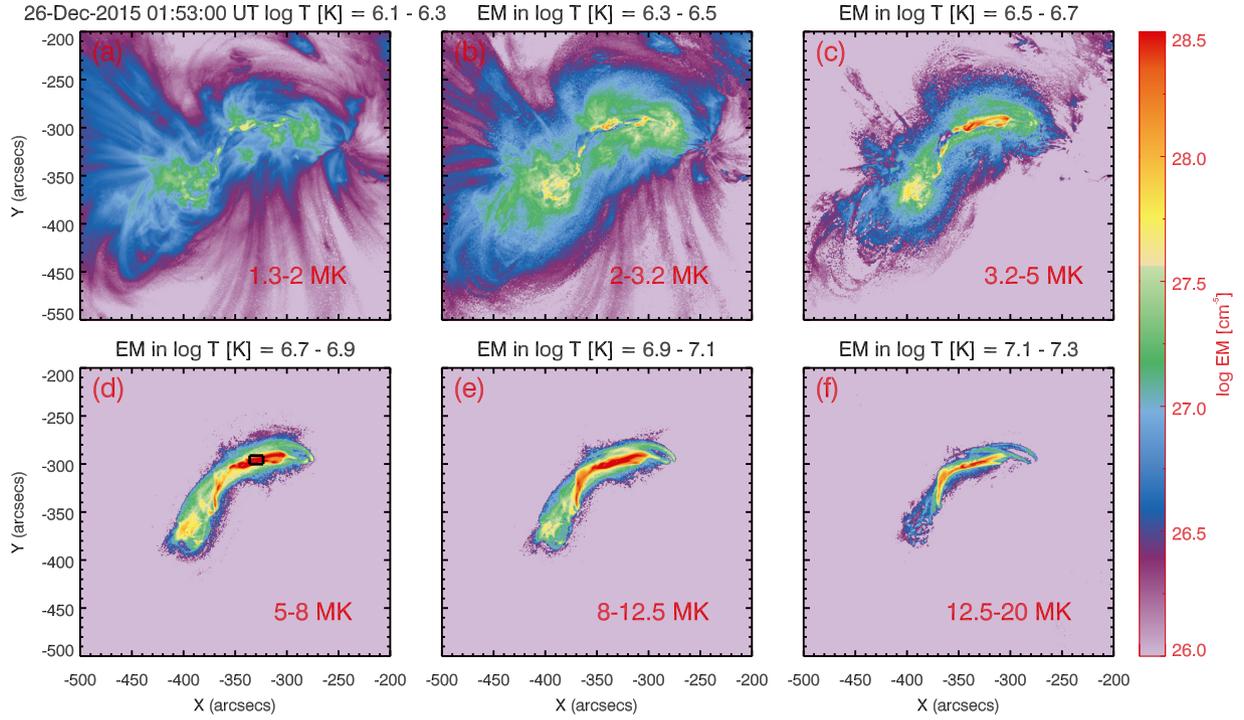

**Figure 6.** EM maps of the sigmoid at the GOES peak time (01:53:00 UT) of the C1.6 X-ray flare. The black boxed regions in the panel (d) is used to plot EM curves in Fig. 8 (see also Movie 3, which is available as supplementary material)

.

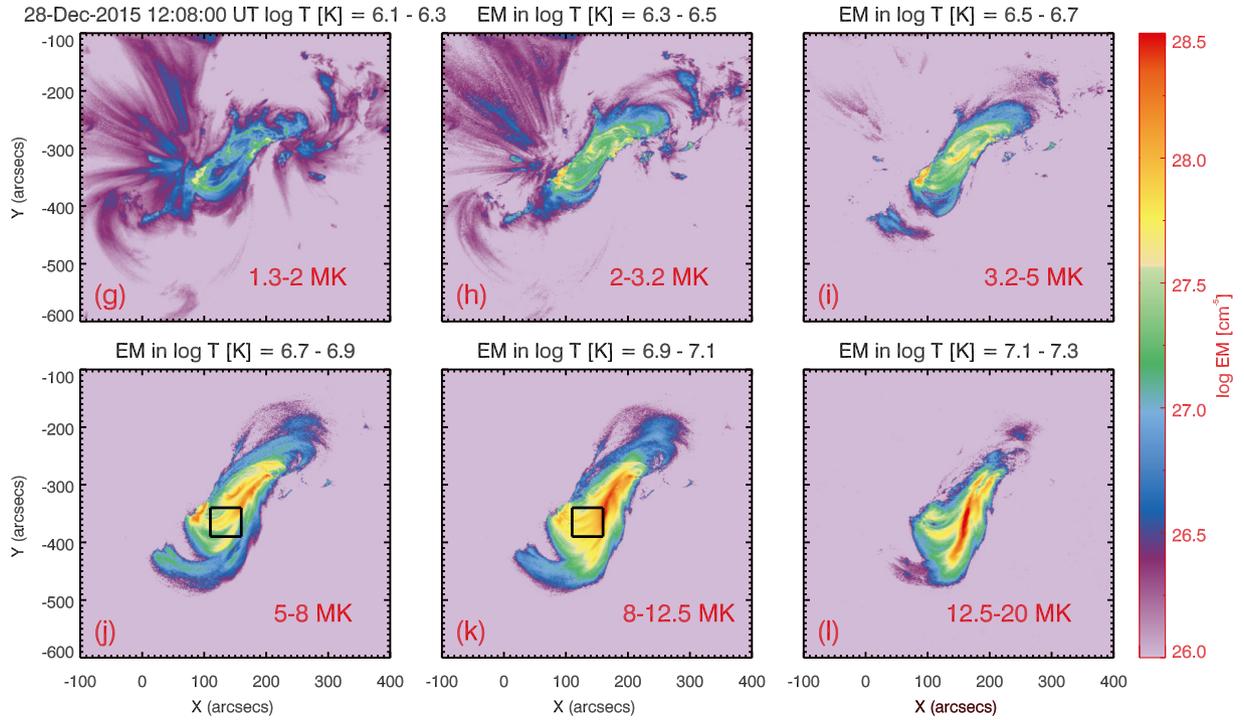

**Figure 7.** Same as Fig. 6 but obtained at the GOES time of peak temperature (12:08:00 UT) of the M1.8 flare. (see also Movie 4, which is available as supplementary material)

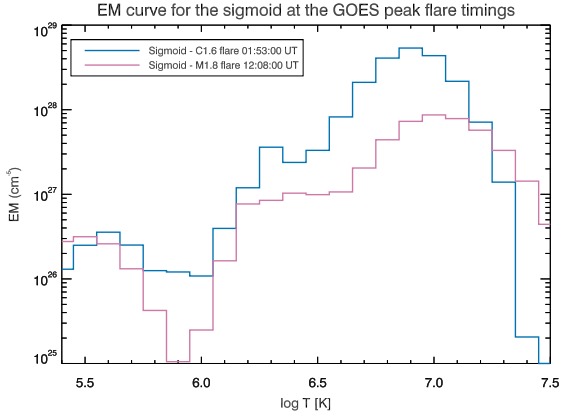

**Figure 8.** Emission measure curves are obtained for the C 1.6 (at the GOES peak time of the flare, 01:53 UT) and the M 1.8 (at the GOES time of peak temperature, 12:08:00 UT) flares. The curves were created for the black boxed regions marked in panels (d) and (j) of Figs. 6 and Figs. 7, respectively.

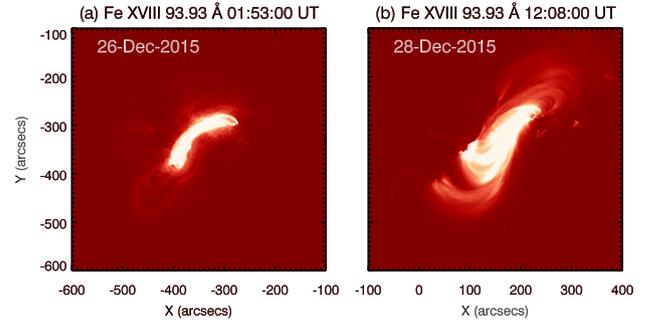

**Figure 9.** Fe XVIII emission derived from the AIA 94 Å channel at the GOES peak time of the C1.6 flare (panel (a)) and at the time of peak temperature of the GOES for the M1.8 flare (panel (b)) flare.

peak temperature of 12.5 MK, which is higher than the GOES peak temperature 10 MK.

An EM analysis was also performed during the M1.8 flare (see also Movie 4, which is available as supplementary material). We display the EM maps in Fig. 7 obtained at the GOES time of peak temperature (12:08 UT). An animation of EM maps obtained at different temperature bins corresponding to different phases of the flare is presented in Movie 4 (which is available as supplementary material). As can be seen, as in Fig. 6, the sigmoidal region is dominated by the emission in high temperatures, $\log T$ [K] > 6.3 (see panels (h) to (l) in Fig. 7). In addition, there are other structures at lower temperatures that corresponding to warm loops and fan loops as seen in the AIA 171 Å and 211 Å images.

By careful examination of the animation, the first high EM was recorded around 11:36 UT at $\log T$ [K] = 6.7-6.9 (5-8 MK) for one of the sigmoidal strands which was located close to northern footpoint of the sigmoidal loops. Over time, the high EM faded away from the EM map at $\log T$ [K] = 6.7-6.9 (5-8 MK). We note that early on, during 11:51–11:53:36 UT, high EM was seen for $\log T$ [K] = 6.5-6.9 (3-8 MK) at the same locations as the bright Ca II ribbons were observed. Interestingly, the locations of brightenings appearing in panel (f) of Fig. 4 show high EM only for $\log T$ [K] = 6.7-6.9 (5-8 MK). The high temperature of >8 MK with high EM values was measured along one of the sigmoidal strands from 12:03 UT onwards. The panel (l) of Fig. 7 shows plasma at much higher temperature (>12.5 MK) with high EM values which were just confined only along a few strands of the sigmoid. This high temperature is comparable to the GOES peak temperature of ∼15 MK.

To obtain a quantitative estimate of the temperature, we have selected an area corresponding to the highest emission as shown by the black boxed area in panel (d) and panel (j) in Figs. 6 and Figs. 7, respectively. We plot the EM curves in Fig. 8 for the C 1.6 (at the GOES peak time of the flare, 01:53 UT) and the M 1.8 (at the GOES time of peak temperature, 12:08:00 UT) flares. We note that the EM curves peak at the $\log T$ [K] = 6.9 (8 MK) and 7.0 (10 MK) for the C1.6 and M1.8 flares respectively. Within the box region, a higher emission measure ($5.36 \times 10^{28}$ cm$^{-5}$) was derived for the C1.6 flare compared to that of the M1.8 flare ($8.68 \times 10^{27}$ cm$^{-5}$).

High temperature plasma above >10 MK was present in both flares and both EM curves fall sharply above $\log T$ [K] = 7.2 (12 MK). Note that the constraints in the high temperature range are limited. For both flares, two lower temperature peaks were recorded at $\log T$ [K] = 6.3 (2 MK) and $\log T$ [K] = 5.6-5.7 (0.4-0.5 MK) indicating contributions from low temperatures plasma. We note that the temperatures corresponding to the peak EMs are lower than those estimated using the GOES-15 observations.

To check the reliability of the temperature obtained using an EM analysis, in Fig. 9, we derived and display Fe XVIII emission from the AIA 94 Å channel at the GOES peak time of the C1.6 flare (panel (a)) and at the time of peak temperature of GOES for the M1.8 flare (panel (b)) flare. The observed structures in the Fe XVIII images match very well with the EM maps obtained in the temperature bin $\log T$ [K] = 6.7-6.9 (5–8 MK) and shown in panels (d) and panel (j) of Figs. 6 and 7. This confirms the high temperature structures (peak formation temperature of ∼7 MK for Fe XVIII emission) observed along the sigmoid.

### 3.3 Temperature estimate using filter-ratio of AIA

Polito et al. (2016, 2017) have demonstrated that, although the AIA channels are multi-thermal, the intensity ratio of a combination of AIA channels can be reliably used as temperature diagnostics for solar flare events. During flares, the AIA 94 and 131 Å channels are dominated by high-temperature lines, Fe XVIII (93.93 Å formed at ∼7 MK) and Fe XXI (128.75 Å formed at ∼10 MK), respectively (see, e.g. Petkaki et al. 2012; Del Zanna 2013; Dudík et al. 2014; Aparna & Tripathi 2016). Under the assumption of an iso-thermal temperature, during flares, the ratio of the AIA 94 and 131 Å channels can be used for temperature diagnostics (see panel (b) of Fig. A1 in Appendix). To get the temperature responses for the AIA 94 and 131 Å channels (see panel (a) of Fig. A1 in Appendix), we followed the same procedure as that outlined by Del Zanna et al. (2011b) using the CHIANTI atomic database (Dere et al. 1997; Del Zanna et al. 2015). An electron number density of $1 \times 10^{10}$ cm$^{-3}$ and elemental coronal abundances as that of Feldman (1992) were used in this analysis.

For this purpose, we used near-simultaneous images of the AIA 94 and 131 Å channels corresponding to GOES flare peak temperature. The temperature maps obtained for the C1.6 and M1.8 flares are shown in the top and bottom panels in Fig. 10, respectively.

During the C1.6 flare (top panel in Fig. 10), it is observed that the plasma along the sigmoid structure was multi-thermal with multiple strands emitting at different temperatures. The plasma temperature went up to ∼10 MK at the flare peak and decreased to 8 MK during the decay phase around 02:25 UT. These temperatures were found to be close with those derived with GOES X-ray fluxes but lower than those from the EM analysis. The peak temperature of $\log T$ [K] = 6.9 (8 MK) from EM curves shows good agreement with the temperature from AIA filter-ratio (panel (a)). During the decay phase (panel (c)) at 02:24 UT (31 minutes after the GOES peak), the temperature of the sigmoid was estimated to be between log $T$ [K] = 6.85 and 6.95 (8-9 MK).

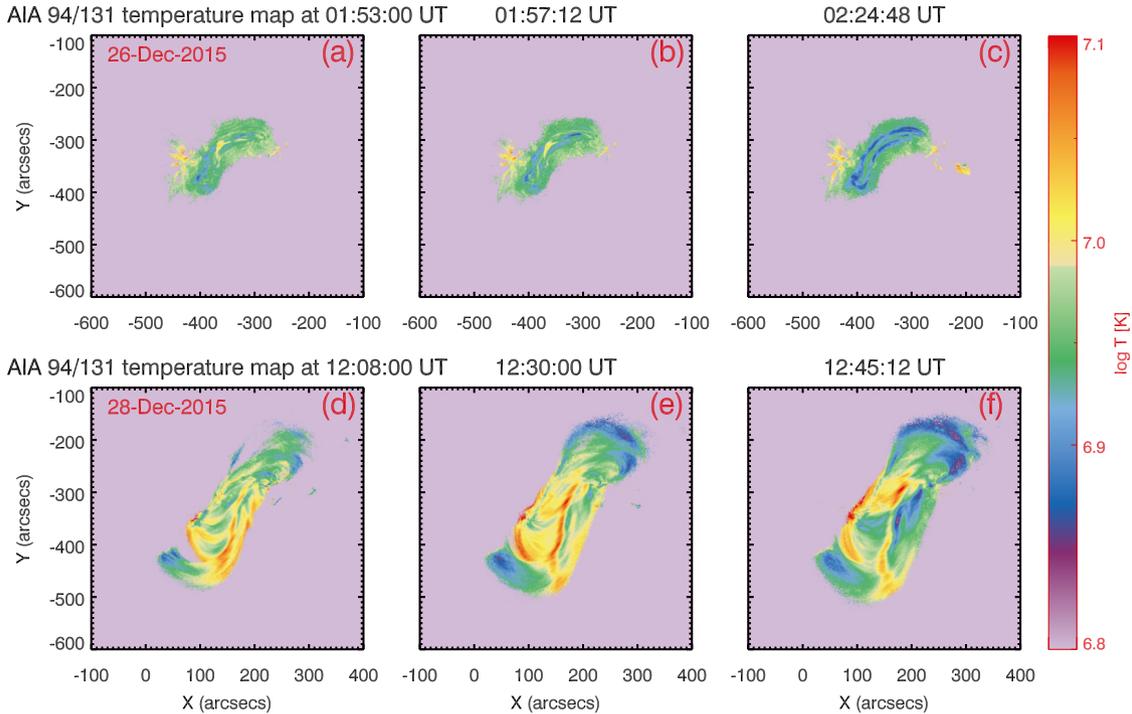

**Figure 10.** The temperature maps obtained using the filter ratio of AIA 94/131 for C-class flare (top row) and M-class flare (bottom row).

The temperature maps during the M1.8 X-ray flare are shown in the panels (d)-(f) of Fig. 10. Firstly, we plotted the AIA filter ratio temperature at the GOES time of peak temperature i.e. at 12:08 UT (see panel (d)). The AIA filter ratio map reveals distinct loops system associated with the sigmoidal structure. A large part of the sigmoid region (yellow patch between X = 50″ to 150″ and Y = -430″ to -360″) was at temperature >10 MK whereas the footpoints of sigmoidal loops were ~8-9 MK. With time, the temperature of that yellow patch region reduced to log $T$ [K] = 6.95 (9 MK) (see green patch area between X = 50″ to 150″ and Y = -430″ to -360″ in panel (e)) at 12:45 UT i.e. the GOES flare peak in 1.0-8.0 Å channel. There is not much change in temperature at the footpoints of the sigmoidal loops at these later times. During the decay phase, the temperature starts to decrease <10 MK in most of the sigmoid, which is in close agreement with the temperature recorded by GOES at that time.

We note that the temperature maps obtained using filter ratios show an excellent match with those obtained using the EM analysis, albeit some of the strands along the sigmoid showed slightly higher temperatures in the filter-ratio maps. However, these temperatures were lower than those obtained from GOES. We attribute this to the fact that GOES derived temperature is an average of the entire solar disk, however, those from EM analysis and filter ratio of images are for localized region on the Sun.

### 3.4 Temperature estimate using X-ray imaging observations

The XRT observations were only available during the C1.6 flare and the average temperature was obtained using the filter-ratio analysis on the unsaturated Al-poly and Be-thin filter images. In Fig. 11, we display the sigmoid recorded by XRT using Al-poly (panels (a), (d), and (g)) and Be-thin filters (panels (b), (e), and (h)). We observe a very similar morphology for the sigmoidal structure, as well as the brightening at the same location as seen in the AIA 94 Å channel.

We obtained XRT temperature maps during the impulsive rise, peak and decay phase of the C1.6 X-ray flare. We estimated temperatures only for unsaturated XRT images, which are not exactly at the same time as EM maps and AIA filter-ratio maps shown in Figs. 6, 7 and 10, but are taken nearly simultaneously. The temperature maps are shown in the panels (c), (f) and (i) of Fig. 11.

We note that during the impulsive phase of the flare, the XRT filters were saturated. The last unsaturated frames before the peak of the flare were taken at 01:44:35 UT in Al-poly and 01:44:47 UT in Be-thin, which are 12 sec apart. The derived temperature map is shown in panel (c) of Fig. 11. During this phase, the sigmoid shows a temperature of about log $T$ [K] = 6.5-6.6 (3.2-4 MK). This is in agreement with the temperatures obtained using the EM analysis (see the EM map at 01:44:36 UT in animation Movie 3, which is available as supplementary material) as well as with the GOES temperature analysis. The unsaturated XRT images were only available ~4 minutes after the peak of the flare. The temperature maps obtained using these unsaturated observations are shown in panels (f) and (i) of Fig. 11. These maps show the presence of plasma at a temperature of about log $T$ [K] = 6.9-7.0 (8-10 MK) along the sigmoidal region, which is in good agreement with the temperature obtained from EM maps, the AIA filter-ratio as well as the GOES temperature.

During the decay phase of the flare, the sigmoid started to cool down from temperature log $T$ [K] = 7.1 to 6.9 (12.5-8 MK), which is in good agreement with the temperatures obtained from AIA EM analysis, the EM maps at 02:06:24 UT (see also Movie 3, which is available as supplementary material) but slightly higher than those estimated from GOES.

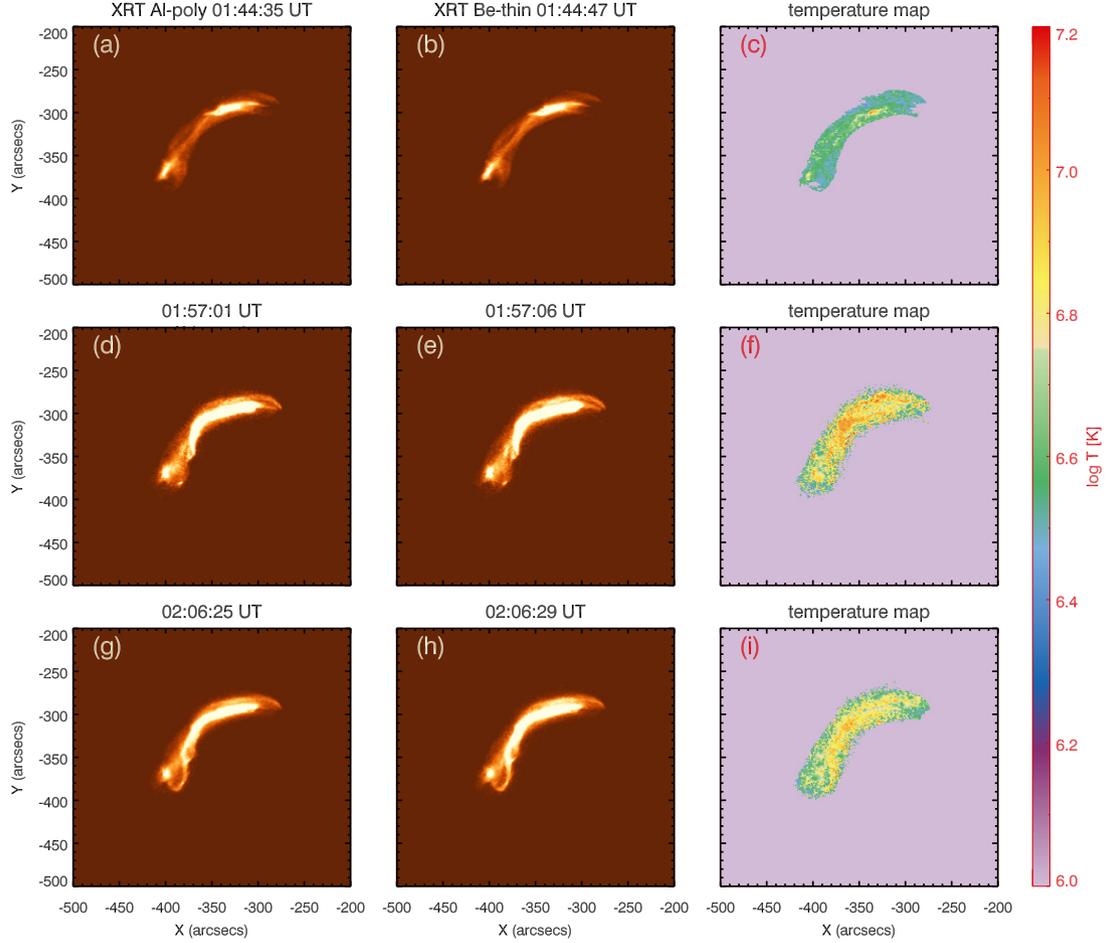

**Figure 11.** XRT Al-poly (panels (a), (d), and (g)) and Be-thin filter (panels (b), (e), and (h)) images of the sigmoid during the C1.6 X-ray flare, respectively. Panels (c), (f), and (i) : the temperature maps of the sigmoid obtained at different times from the intensity ratio of the two XRT channels using the XRT filter-ratio method.

### 3.5 Temperature analysis for all flares

There were three big sunspots associated with NOAA active region #12473 on Dec. 24 which showed $\beta\gamma$ magnetic configuration. With passing time, the active region magnetic field became more complex and showed $\beta\delta$ configuration before the final sigmoid eruption on Dec. 28. As stated earlier, there were 16 X-ray flares (four B-class, ten C-class, and two M-class) observed during the lifetime of the sigmoid on the solar disk from Dec. 24 to 28, 2015. The sigmoid produced one M-class flare on Dec. 24, six C-class flares on Dec. 25, four C-class and one B-class flare on Dec. 26, three B-class flares on Dec. 27 and one M-class flare during the sigmoid eruption on Dec. 28. Because of the low signal-to-noise ratio in GOES, AIA, and XRT, we had discarded all B-class flares, which originated from the sigmoid and studied the temperature structure of the sigmoid during all C and M class flares. The GOES X-ray fluxes along with Fe XVIII maps at the pre-flare and peak timings of the flares are available as supplementary information.

A similar analysis as given in sections 2 and 3 was carried out for all the C and M class flares and details are given in Table A. In column 1, we note the date of the observation, NOAA active region number, its heliographic coordinates, and configuration of the active region related to sunspots. We indicate flares with numbers in the order of their occurrence (see column 2). Columns 3-6 indicate the class of the GOES X-ray flare observed in the 1.0-8.0 Å channel, phase of the flare, the time related to different phases of the flare obtained from NOAA and the temperature obtained using two GOES X-ray channels, respectively. The unsaturated AIA images at the same time given in column 5 were used to obtain temperatures using an EM analysis. The range of temperatures where the sigmoid region was observed in the EM maps are noted in column 7. Further, we subtracted EM maps (with the same temperature intervals as given in Fig. 6) obtained at the pre-flare timings from those EM maps which were obtained at the GOES start, peak and end timings. The average EM for sigmoidal region was obtained for those new subtracted EM maps. The temperature intervals for highest averaged EM are given in the parenthesis in column 7.

The temperatures during the peak and decay phase of the flares using the AIA filter-ratio method were obtained at the same timings as given in column 5. The range of temperatures where the sigmoid is observed are given in column 8. The average temperatures of the sigmoidal region from the AIA filter ratio maps were obtained and are given in column 8 in parenthesis. The blank spaces (marked as horizontal dashed lines) for the peak and end of the flares in column 8 indicate that the temperatures could not be estimated because of saturated AIA images. The temperatures of the sigmoid during flares were obtained by employing the XRT filter-ratio method on the near-simultaneous XRT images (with respect to the start, peak and end times of GOES X-ray flares). The time for XRT images and XRT temperatures are mentioned in columns 9 and 10, respectively. The average temperature over the sigmoidal region was obtained from the XRT temperature map and the temperatures are given in column 10 in parentheses. The coronal and chromospheric activity during each flare is briefly described in the Appendix A.

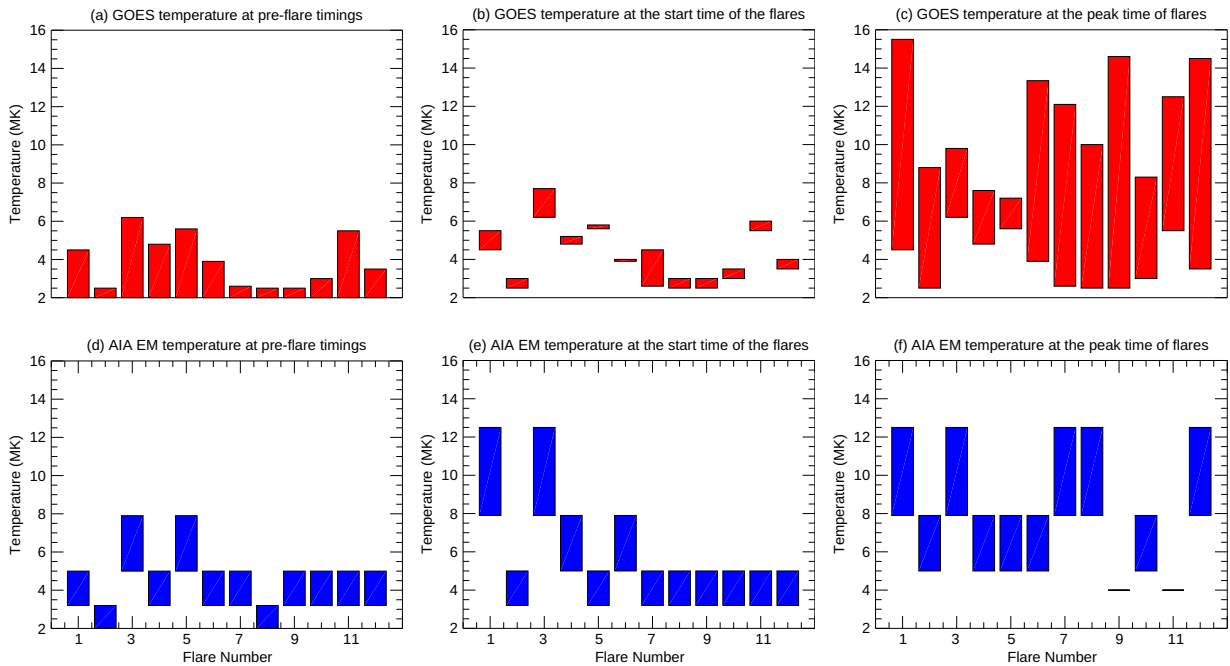

**Figure 12.** Bar plot of temperatures obtained from the GOES channels (red), and peak in the EM analysis (blue) at the pre-flare timings (left column), at the start (middle column) and peak (right column) times of the flares. The blank spaces in temperatures indicate that the AIA data are saturated.

A pictorial representation of the temperature information given in Table A is shown in Fig. 12. The temperature ranges for all flares obtained from the GOES channels and AIA EM analysis (those given in Table A, Column 7, parentheses) at the pre-flare, start and peak timings of the GOES flare are plotted as bar plots. We assumed that the initial temperature recorded by GOES is 2 MK before the temperature noted for pre-flare GOES timings and plotted this as the lower end for each bar in panel (a). The temperature obtained at the start time of the flare (i.e. upper limit for each bar in panel (a)) was taken as a starting point (lower limit) for each bar plotted in panel (b) and (c). The higher temperatures for each bar in panel (b) indicate temperatures obtained at the start time of the flare whereas in panel (c), the higher temperatures for each bar indicate the temperature obtained at the peak of the flare. In the bottom panel ((d), (e) and (f)) of Fig. 12, we plotted the temperature ranges given in the parenthesis of column 7 which were obtained at the highest averaged EM over the sigmoidal region. The consistent rise in temperature of the sigmoid was observed from the pre-flare phase time to the peak time of the flare in both GOES and AIA EM analysis.

Even though the individual pixels (and strands along sigmoid) in the XRT temperature maps show temperatures ranging from 4–10 MK, the averaged XRT temperatures (those given in Table A, Column 10, parentheses) of the sigmoid were found to be lower for all flares. This, and the sparsity of XRT data, has restricted us from comparing the averaged XRT temperatures with temperatures obtained from the GOES observations (in panel (a), (b) and (c)) and AIA EM analysis (panel (d), (e) and (f)).

Based on the temperature analysis of all flares (see Table A), we obtained the following results. We compared the GOES temperatures obtained for all 12 flares with those obtained using the EM analysis, AIA and XRT filter-ratios. The lowest (highest) temperature of 7.2 MK (15.5 MK) was measured at GOES peak time of C1.1 (M1.1) flares on Dec. 25 (Dec. 24) whereas the EM analysis showed the sigmoid structure was at ∼3 MK at the start of all flares and reached a high temperature of ∼12.5 MK for all flares except for the C1.0 and C1.1 flares on Dec. 25 where the temperature was ∼10 MK. The AIA filter ratio recorded a temperature range of 8–12.5 MK at the peak time of the flares, whereas the temperature of 4–10 MK was obtained from the XRT filter-ratio. Sterling et al. (2000) reported a pre-flare sigmoid temperature of 2.4 MK using the filter-ratio method. This found to be lower than the XRT temperatures we obtained for all flares we studied.

In the statistical study of flares by Feldman et al. (1996), the authors obtained a linear relationship between the electron temperature that was obtained at the peak of the flares using the Bragg Crystal Spectrometer (BCS) onboard Yohkoh and the class of the X-ray flares recorded by GOES. They obtained a range of temperatures of 7-14 MK for GOES C-class flares and 13-22 MK for GOES M-class flares. In our analysis, the peak temperatures of 10–12.5 MK obtained from all four methods are found to be within the range of temperatures provided by Feldman et al. (1996) for C and M-class flares.

## 4 SUMMARY AND CONCLUSIONS

In this paper, we have carried out a thorough investigation of the thermodynamic nature of a flaring sigmoid structure. The main aim of the study is to determine the distribution of temperatures along sigmoid structures during different phases of the flares. By making use of AIA and XRT imaging observations as well as X-ray fluxes from the GOES, we obtained reliable estimates of temperatures along a sigmoid using the emission measure analysis and filter-ratio methods. The following are the main results that we obtained from our analysis.

- The sigmoid activity started with a brightening (marked as a blue star in panel (a) of Fig. 3) in the AIA 94 and 131 Å channels followed by chromospheric brightenings in AIA 1700 Å and SOT Ca II images (see Figs. 3 and 4).

- The temporal and spatial correlation between the intensity variation along the sigmoid in the AIA channels and GOES X-ray flares was confirmed (see Fig. 5).

- A systematic delay in the intensity peak of the AIA (94 and 131 Å) and GOES channels indicate that there was an increase in the temperature of the sigmoid during the rise phase of flares. This was confirmed using various temperature diagnostic techniques.

- The sigmoid structure was seen to be multi-thermal (3-12.5 MK) throughout flaring activity. Some strands along the sigmoid were observed at a high temperatures ~12.5 MK or greater at the peak time of the flare.

- We confirmed that emission from high temperature Fe XVIII and Fe XXI lines was present at the sigmoid location at the peak time of the flare. In some flares, this hot emission was confined to some strands which were part of the sigmoid and in some flares, the hot emission spread over the whole sigmoid.

The AIA and XRT observations confirmed the presence of S-shaped elongated loop structures associated with a sigmoid which formed along the PIL. The variation in EUV intensities along the sigmoid and GOES X-ray fluxes confirmed that the flaring activity occurred along the sigmoid (see also Movie 1 and 2, which is available as supplementary material). The small initial chromospheric brightenings coincided with the AIA brightening in the coronal channels. This indicates that the energy dissipation in the chromosphere occurred at the same locations as the coronal brightenings. In addition, two bright ribbons were observed at the footpoints of sigmoidal loops during the M1.8 flare, and with time, they started to move away from each other (away from the PIL). This motion results in the strong shearing motion in the sigmoidal loops and eventually, the process most likely led to the eruption of the sigmoid. A new bundle of elongated loops (post flare loops) started to form immediately at the same sigmoidal location. A footpoint drifting motion has also been observed in another M-class flare studied by Zemanová et al. (2019). These authors interpreted the hot sigmoid loop as a flux rope using a three-dimensional (3D) extension to the standard solar flare model, originally proposed by Tripathi et al. (2006a).

Our temperature investigation of the M1.8 flare using various techniques led us to confirm that there was a systematic rise in the sigmoid temperature corresponding to a systematic delay seen in the peak intensities of the high temperature AIA as well as GOES channels during M1.8 flare. Before the flares started, the GOES recorded the sigmoid temperature of 2.5 MK and 4 MK and the temperature rose to 10 MK and 15 MK at the GOES peak time of the C1.6 and M1.8 flares respectively. This temperature was achieved in 9 min for C1.6 flare compared to slow rise time of 1 hour 25 min for M1.8 flare. The high temperature of 7.1 MK and above was confirmed at the GOES flares peak times using Fe XVIII emission maps. The analysis also confirms that emission from high temperature Fe XVIII and Fe XXI lines was seen in the AIA 94 and 131 Å channels respectively (Tripathi et al. 2013, Patsourakos et al. 2013, Aparna & Tripathi 2016).

The EM analysis of the sigmoid structure on both days shows the multi-thermal plasma associated with the sigmoidal structure. During the C1.6 flare, the EM maps show a small amount of lower temperature plasma (log $T$ [K] < 6.3) in the active region, whereas the sigmoid was dominated by high-temperature plasma (log $T$ [K] >6.3). Some of the loops along the sigmoid showed high EM as well as high-temperature (10-12.5 MK) for both flares (see panel (e) and (k) of Fig. 6 and 7, respectively). These temperatures were close to those obtained by GOES but slightly higher. Before the start of M1.8 flare, AIA showed a bundle of low-temperature loops ~1-2 MK as well as high-temperature >3 MK loops. A few sigmoidal loops and their footpoints were seen to be at 5-8 MK. This was found to be slightly lower but consistent with the AIA filter-ratio temperature. A few sigmoidal loops reached very high-temperature of 15 MK and above. The temperatures at peak EM recorded for the boxed regions for both flares were lower (8 MK) than those estimated using GOES channels. The assumption of plasma temperatures >5.6 MK using the AIA filter-ratio during the peak and gradual decay of both flares was found to be correct and provided consistency in the plasma temperature compared with the GOES temperature as well as the temperatures obtained from the EM analysis and XRT filter ratio.

A detailed study of the temperatures of a sigmoid during various phases of solar flares is given in Section 3.5 and Appendix A. The EM analysis confirmed the presence of hot plasma of temperature around 10-12.5 MK in strands of the sigmoidal structure obtained at the peak timings of all flares. The temperature range of 8–12.5 MK from AIA filter ratio and 4–10 MK from XRT filter-ratio supports the notion that Fe XVIII emission (available as supplementary information) formed at a temperature corresponding to the peak of the contribution function of Fe XVIII line. Knowing the constraints on the EM analysis, simultaneous spectroscopic observations of flaring sigmoid would be useful to see whether the sigmoid can reach to a temperature higher than 12.5 MK at the peak time of flares.

Using spectroscopic observation from CDS, Gibson et al. (1999) and Tripathi et al. (2006b) reported a maximum temperature of 2 MK for the quiescent sigmoid that was observed using Fe XVI and Si XII lines. This temperature is lower compared to those obtained in this paper and is close to the pre-flare sigmoid temperatures. Del Zanna et al. (2002) reported the presence of Fe XIX emission along a sigmoid and found the temperature of 8 MK, whereas Cheng et al. (2014) and James et al. (2018) reported higher temperatures, 10 MK and 7-14 MK, respectively. These results are in good agreement with the temperatures we obtained during the sigmoidal flaring events using the EM analysis.

The temperatures obtained using GOES, EM analysis and filter-ratio methods show overall consistency for flaring sigmoidal. Together, these results provide important insights into our understanding of the thermal structure of sigmoids and are important ingredients for the thermodynamics modelling of the sigmoidal structures and associated flares in the core of active regions. This would also help to the design of simultaneous imaging and spectroscopic observations using existing and upcoming instruments. The spectroscopic observations will facilitate to measure plasma parameters such as electron number density, plane-of-sky, Doppler velocities and Non-thermal velocities at multiple temperatures. It is interesting to note here that with increasing complexity in the magnetic field of sunspots which were associated with sigmoidal region, a number of the flares as well as the eruption occurred at the similar location shown as blue star in Fig. 3. The time evolution study of the photospheric magnetic field may provide important insights into understanding the cause for these flares and the eruption, possibly giving an indicator for sigmoidal instability. The MHD modelling as well as NLFFF using vector magnetogram (analysis similar to that reported by Jiang et al. (2013, 2014, 2018)) is necessary to further confirm the location of the null point, bald patch separatrix surface, and formation of a torus instability which result in enhanced heating of the sigmoid and its eventually eruption.

## ORCID ID'S

Sargam M. Mulay  https://orcid.org/0000-0002-9242-2643

Durgesh Tripathi  https://orcid.org/0000-0003-1689-6254

Helen Mason  https://orcid.org/0000-0002-6418-7914

## ACKNOWLEDGEMENTS

Part of this work was carried out when SMM held a postdoc position at IUCAA. Currently, SMM is a research assistant at the University of Glasgow and acknowledges support from the UK Research and Innovation's Science and Technology Facilities Council under grant award numbers ST/P000533/1 and ST/T000422/1. This research is partly supported by the Max Planck partner group on the *Coupling and Dynamics of the Solar Atmosphere* at IUCAA. HEM thanks the UK's Science and Technology Facility Council for financial support, the Cambridge/India Hamied Visiting Lecture Scheme for travel funds and IUCAA for hosting her visits. The authors also thank Dr. Sreejith Padinhatteeri from IUCAA for useful suggestions and comments. AIA data are courtesy of SDO (NASA) and the AIA consortium. The GOES 15 X-ray data are produced in real time by the NOAA Space Weather Prediction Center (SWPC) and the data are distributed by the NOAA National Geophysical Data Center (NGDC). Hinode is a Japanese mission developed and launched by ISAS/JAXA, collaborating with NAOJ as a domestic partner, NASA and STFC (UK) as international partners. Scientific operation of the Hinode mission is conducted by the Hinode science team organized at ISAS/JAXA. This team mainly consists of scientists from institutes in the partner countries. Support for the post-launch operation is provided by JAXA and NAOJ (Japan), STFC (U.K.), NASA, ESA, and NSC (Norway). CHIANTI is a collaborative project involving George Mason University, the University of Michigan (USA), and the University of Cambridge (UK). NOAA Solar Region Summary data supplied courtesy of SolarMonitor.org

## DATA AVAILABILITY

In this paper, we used the Interactive Data Language (IDL) and SolarSoftWare (SSW; Freeland & Handy 1998) packages to analyse GOES, AIA, HMI and SOT data. Some of the figures within this paper were produced using IDL colour-blind-friendly colour tables (see Wright 2017). The AIA and HMI data is available at http://jsoc.stanford.edu/ and the data were analysed using routines available at https://www.lmsal.com/sdodocs/doc/dcur/SDOD0060.zip/zip/entry/. The SOT data was obtained from archive available at http://sdc.uio.no/search/form and data analysed using routines available at https://sot.lmsal.com/Data_new.html#using_data. The solar flare details are obtained from the archive ftp://ftp.swpc.noaa.gov/pub/warehouse/.The GOES data analysis was performed by following IDL routines available at https://hesperia.gsfc.nasa.gov/rhessidatacenter/complementary_data/goes.html

# APPENDIX A:  ADDITIONAL MATERIAL

## AIA TEMPERATURE RESPONSES

We followed the procedure given in the Appendix of Del Zanna et al. (2011b) to obtain the temperature response function for the AIA 94 and 131 Å channels. The method uses the SSW routine `aia_get_response.pro` to calculate wavelength responses (effective areas) by considering the time-dependent instrument degradation. An isothermal spectrum was obtained using the CHIANTI atomic database (Dere et al. 1997, Del Zanna et al. 2015) including all the lines and continuum over the wavelength range from 90 to 140 Å. The temperature responses were obtained by convolving isothermal spectra with effective areas. An electron number density of $1 \times 10^{10}$ cm$^{-3}$ and elemental coronal abundances by Feldman (1992) were used in this analysis. The temperature response functions for these channels are shown in panel (a) of Fig. A1 and the ratio is shown in panel (b). There are two peaks in responses at lower and higher temperatures for both channels, hence the AIA 94/131 ratio is not a single valued function.

In the case of the flare peak and during decay phase, we assumed that the average plasma temperature along LOS is above log $T$ [K] = 6.75 (5.6 MK). The ratio of observed intensities in the AIA 94 and 131 Å channels were measured and compared with this theoretical ratio and the averaged temperature was estimated (see section 3.3).

## XRT TEMPERATURE RESPONSES

Using the filter-ratio method (See section 5 of Narukage et al. 2011), near-simultaneous images in the XRT filters could be used for coronal temperature diagnostics (Polito et al. 2016). Care should be taken while interpreting the temperature if there is a large time difference in two filter images and/or the images are saturated which may result in under or over estimation of the plasma temperature. In a similar way to the AIA responses, we calculated the temperature responses for Al-poly and Be-thin filters by following the procedure given by Del Zanna et al. (2011b). The effective areas and spectral responses were obtained using the `make_xrt_wave_resp.pro` routine. An electron number density of $1 \times 10^{10}$ cm$^{-3}$ and elemental coronal abundances by Feldman (1992) were used in this analysis. The effective areas were then folded with the isothermal spectrum and XRT temperature responses were obtained.

The temperature responses in both channels are shown in panel (c) and the ratio is shown in panel (d) of Fig. A1. The XRT filters are broadband and are sensitive to high-temperatures >2.5 MK. The temperature response function for each XRT filter is single-valued with only one peak, unlike AIA channels. Therefore, the ratio of temperature response of two XRT filters is a single-valued function. The ratio of observed intensities in both XRT channels were measured and compared with this theoretical ratio and averaged temperature was estimated (see section 3.4).

## X-RAY FLARE #1: 24-DEC-2015 - M1.1 CLASS

The AIA images on Dec. 24, showed that the active region #12473 consisted of three sunspots with a magnetic configuration of $\beta\gamma$ and was closed to the solar East limb. Because of its close proximity to the limb, the S-shape of the sigmoid structure was not obvious. The GOES observed an M1.1 X-ray flare on Dec. 24 at 01:49 UT and the X-ray fluxes showed two peaks; a small peak during the rising phase of the flare just before the actual peak of the flare. In GOES 0.5-4.0 Å channel, the first peak was observed at 01:57 UT and the second peak was at 02:10 UT whereas in GOES 1.0-8.0 Å channel, the first peak was observed at 02:00 UT and the second peak was at 02:12 UT. These timings indicate the delay of ~2 min in both the peaks of two GOES channels. The GOES recorded temperatures at the pre-flare of 4.5 MK at 01:44 UT. The temperature began to rise from 5.5 MK at 01:49 UT and showed a first peak temperature, 15.5 MK at 01:58 UT. The temperature began to decrease and reached up to 12.5 MK at 02:04 UT. It started to rise again and showed a second peak temperature, 15.5 MK at 02:10 UT. The plasma temperature started to decrease during the decay phase and recorded a temperature of 11 MK at 02:22 UT. An intense brightening was observed along the sigmoid in all AIA channels and most of the channels got saturated during the rising phase of the flare i.e. before the second peak. Because of the unavailability of SOT Ca II data, the chromospheric brightening was only observed in the AIA 1600 and 1700 Å channels during the flare. The chromospheric brightening was observed along the PIL.

The AIA 94 Å channel showed the presence of Fe XVIII emission at the pre-flare phase and also at different phases of the flare. The EM maps showed the presence of multithermal plasma along the sigmoid. The pre-flare temperature of 3-5 MK was recorded but there was a small amount of plasma observed at higher temperatures 6-8 MK. The temperature as well as EM increases during the flare. The maximum temperature between 10-12.5 MK was recorded at 02:06 UT. We could not determine temperatures during the peak because of saturated AIA images. The AIA filter-ratio recorded similar temperatures, 10-12.5 MK at 02:06 UT and shows good agreement with EM peak temperature. The GOES peak temperature were higher than those obtained by the EM analysis and AIA-filter ratio. There were no X-ray observations available during the flare.

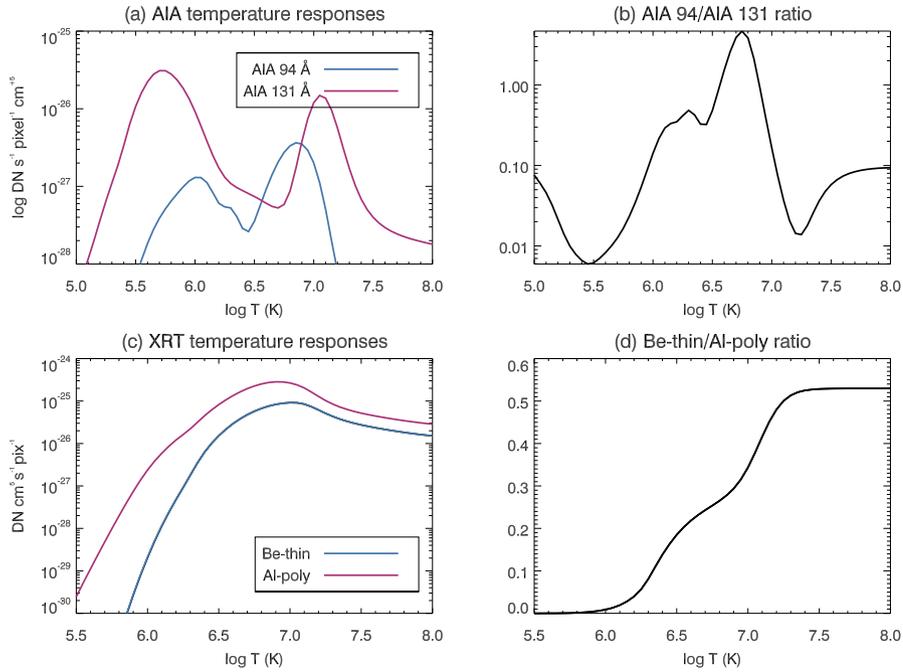

**Figure A1.** Panel (a) and (c): The temperature responses for AIA and XRT channels. Panels (b) and (d): the filter-ratios for AIA and XRT.

**X-RAY FLARE #2: 25-DEC-2015 - C 1.5**

The S-shaped coronal loops associated with the sigmoid are nicely observed in all AIA channels. The hotter AIA channels show an intense brightening along the sigmoid during the flare. The GOES observed a C 1.5 X-ray flare on Dec. 25 at 04:05 UT. In a similar way to the M 1.1 flare observed on Dec. 24, this flare showed a small peak in the rising phase. The GOES observed 2.5 MK plasma before the flare starts and recorded 3.0 MK and 8.8 MK plasma temperature during the rise and at the peak of the flare respectively. The SOT missed the start of the flare and observed the chromosphere in Ca II from 04:09 UT. The chromospheric signatures were not as strong as other flares. Small brightenings were seen along PIL.

Fe XVIII emission was confined at a very small location along the sigmoid. Very close to the peak of the flare (from 05:12 UT onwards), the sigmoid loops start to fill with Fe XVIII emission and spread over the entire sigmoid at the peak of the flare. The EM maps showed multithermal plasma along the sigmoid. The entire sigmoid structure was mostly dominated by plasma above 2 MK and it reached temperature 10-12.5 MK during the peak of the flare. These temperatures are slightly higher than those recorded by AIA and XRT filter-ratio whereas the GOES peak temperature was lower than the temperatures obtained by all other methods.

**X-RAY FLARE #3: 25-DEC-2015 - C 2.0**

The AIA images showed more elongated loops along the sigmoid. Also a similar brightening was observed in all channels. The GOES observed a C 2.0 X-ray flare on Dec. 25 at 09:20 UT and recorded temperatures at the pre-flare (6.2 MK at 09:15 UT), rise (7.65 MK at 09:20 UT), peak (9.76 MK at 09:38 UT) and decay (7 MK at 10:01 UT) phases. The chromospheric brightenings were confined to small locations at the footpoints of the sigmoidal structure where it was anchored in the sunspots. Fe XVIII emission along the sigmoid started to be seen after 09:20 UT and filled the entire sigmoid by the peak of the flare.

The EM maps showed multithermal plasma along the sigmoid which was mostly dominated by plasma temperature >2 MK. There were loops seen to be at temperature 3-5 MK whereas the hot plasma >5 MK was only discernible along a few core loops. The EM maps showed 10-12.5 MK plasma at the peak of the flare with high EM values for some loops. These results are higher than the temperatures obtained with GOES, AIA and XRT filter ratio. Before the flare started, there were some hot loops associated with the active region but they were not a part of the sigmoid structure. We could not measure temperature from X-ray images because of the saturated pixels along the sigmoidal structure. But during the peak, the XRT recorded the temperature of 5-10 MK which further decreased to 5-9 MK during the decay phase. The XRT peak temperature is close to the temperature obtained from the AIA filter ratio but higher than GOES peak temperature.

**X-RAY FLARE #4: 25-DEC-2015 - C 1.0**

The S-shaped sigmoid showed a brightening in all AIA channels during the C 1.0 X-ray flare on Dec. 25 at 12:33 UT. The GOES recorded temperatures at the pre-flare (4.8 MK at 12:28 UT), impulsive rise (5.2 MK at 12:33 UT), peak (7.6 MK at 12:45 UT) and decay (5 MK at 13:04 UT) phases. No significant chromospheric signatures were seen during the flare. Fe xviii emission along the sigmoid was present throughout the flare evolution.

The EM maps showed multithermal plasma along the sigmoid and reached a plasma temperature of 10 MK during the flare. This temperature is similar to AIA and XRT filter ratio temperatures but slightly higher than the GOES peak temperature.

### X-RAY FLARE #5: 25-DEC-2015 - C 1.1

The sigmoid structure remained the same but showed a small enhancement in the intensity during the flare. The GOES observed C 1.1 X-ray flare on Dec. 25 at 13:07 UT and recorded temperatures at the pre-flare (5.6 MK at 13:02 UT), impulsive rise (5.8 MK at 13:07 UT), peak (7.2 MK at 13:14 UT) and decay (4.9 MK at 13:25 UT) phases. No significant chromospheric activity was observed except a small brightening (at the start of the flare at 13:07 UT) at the northern end of the sigmoid. Fe xviii emission along the sigmoid was present throughout the flare evolution.

The sigmoidal loops were multithermal and most of the loops showed temperatures 1-5 MK. During the peak of the flare, the EM maps showed a plasma of temperature of 10 MK along sigmoid. This temperature found to be higher than the GOES and AIA filter ratio temperatures but similar to XRT filter-ratio temperature. The GOES peak temperature was lower than temperatures obtained from all other methods.

### X-RAY FLARE #6: 25-DEC-2015 - C 6.9

Between the pre-flare activity (from 16:54 UT) and peak (at 17:19 UT) of the C 6.9 X-ray flare on Dec. 25, the GOES recorded multiple peaks in both GOES channels. During the rising phase (∼22 min), the plasma temperature started to increase from 3.9 MK and reached to 7.5 MK at 17:16 UT before it increases to the peak temperature of 13.3 MK at 17:19 UT. No significant chromospheric activity was observed except small brightenings at the ends of the sigmoidal structure. Similar to other flares, the brightening was observed in all AIA channels, mostly the hotter channels showed intense brightening along sigmoid. The AIA 94 Å images also showed emission from Fe xviii line.

High EM values were obtained at the southern end of the sigmoid whereas EM maps recorded a plasma temperature of 12.5 MK during the peak of the flare which is found to be lower than the GOES temperature but higher than the XRT and AIA filter ratio temperature. There were no XRT data available during pre-flare activity and rising time but the XRT filter-ratio method gave a plasma temperature of 10.0 MK at the peak of the flare.

### X-RAY FLARE #7: 25-DEC-2015 - C 3.0

In a similar way to other flares studied, AIA showed brightenings along the sigmoid during the C 3.0 X-ray flare on Dec. 25 at 22:48 UT. The GOES recorded temperatures at the pre-flare (2.6 MK at 22:43 UT), impulsive rise (4.5 MK at 22:48 UT), peak (12.1 MK at 23:00 UT) and decay (7 MK at 23:14 UT) phases. An intense chromospheric brightening was seen at the northern end of the sigmoid just before the flare peak. As time passed, the brightening was seen to be spreading over the other parts of the sigmoid along the PIL. The AIA 94 Å images also showed emission from the Fe xviii line along the sigmoid.

The EM maps show multithermal plasma and recorded a plasma temperature of 12.5 MK during the peak of the flare. This is found to be consistent with the GOES temperature but higher than the AIA filter ratio temperatures. Using the XRT filter ratio, a plasma temperature of 3–4 MK was obtained before the flare started. There was a data gap for the Al-poly and Be-thin channels for ∼50 min from 22:41 to 23:30 UT, so temperature could not be determined. During the decay phase of the flare, the XRT data was available and recorded a plasma temperature of 4–9 MK.

### X-RAY FLARE #9: 26-DEC-2015 - C 7.8

From the GOES fluxes it is observed that the rising phase of the C 7.8 X-ray flare on Dec. 26, 2015 was not smooth, but showed a couple of small peaks. During the rising phase of 10 min, the plasma temperature increased from 4 MK and reached 13 MK at 05:06 UT. Then it decreased by 1 MK between 05:06 and 05:07 UT before peaking again at 14.6 MK. The chromospheric signatures first observed at 04:58 UT at northern end of the sigmoid. The brightening slowly spread over the PIL and reached the southern end of the sigmoid at 05:08 UT. The Fe xviii emission was observed to spread over sigmoid and after 05:09 UT (close to the peak of the flare), almost all the AIA channels were saturated.

The EM analysis showed the temperature of 10 MK at the flare peak time at 05:06 UT. The XRT recorded the plasma temperature of 3–4 MK during the pre-flare and rising phase. We could not measure temperature at the peak of the flare because of saturated XRT images.

### X-RAY FLARE #10: 26-DEC-2015 - C 1.2

The GOES recorded the C 1.2 X-ray flare on Dec. 26 at 11:55 UT and observed a small peak in the rising phase where the plasma temperature was 8.4 MK at 11:58 UT. The plasma temperature decreased to 7.7 MK at 12:02 UT and then reached to 8.3 MK at 12:03 UT. The chromospheric brightening was only observed at the northern part and end of the sigmoidal structure. The AIA 94 Å images also show emission from Fe xviii line along the sigmoid.

High EM values were measured at the similar location and temperature was 12.5 MK. This temperature was in agreement with XRT filter-ratio temperature but found to be higher than GOES temperature.

### X-RAY FLARE #11: 26-DEC-2015 - C 4.1

The GOES recorded a C 4.1 X-ray flare on Dec. 26 at 15:10 UT. There was a small peak in the rising phase at 15:13 UT before the peak at 15:26 UT. The GOES showed a plasma temperature was 5.5 MK before the flare which reached to 9.8 MK during the rising phase. The plasma temperature then decreased to 8.5 MK at 15:20 MK before reaching another peak temperature of 12.5 MK at 15:26 UT. An intense brightening along the PIL was observed in the Ca ii images. The Fe xviii emission was present along the sigmoid since pre-flare phase and the temperature of the sigmoid increased as the flare progressed.

The temperatures from AIA could not be determined because of the saturated images. Unfortunately, we could not determine temperature at the peak and decay phase of the flare from XRT because there were no observations. The XRT data were only available during the pre-flare and rising phase, where a plasma of temperature of 4–5 MK was derived.

Table A1: Sigmoid and X-ray flare observation details

| (Col. 1) | (Col. 2) | (Col. 3) | (Col. 4) | (Col. 5) | (Col. 6) | (Col. 7) | (Col. 8) | (Col. 9) | (Col. 10) |
|---|---|---|---|---|---|---|---|---|---|
| Date | Flare no. | X-ray Class | Phase of the flare | Time (UT) | GOES temp. (MK) | Temp. from EM (MK) | AIA filter-ratio temp (MK) | Time (UT) | XRT filter-ratio temp (MK) |
| 24-Dec-2015 AR #12473 S23 E36 (βγ) | 1 | M1.1 | pre-flare start | 01:44:00 | 4.5 | 2.0-8.0 (3.2-5.0) | - | No XRT data | - |
| | | | start | 01:49:00 | 5.5 | 2.0-8.0 (7.9-12.5) | - | | - |
| | | | peak | 02:12:00 | 15.5 | 2.0-12.5 (7.9-12.5) | 10.0-12.5 (9.87) | | - |
| | | | end | 02:22:00 | 11.0 | 2.0-8.0 (5.0-7.9) | 8.0-11.0 (9.79) | | - |
| 25-Dec-2015 AR #12473 S22 E22 (βγ) | 2 | C1.5 | pre-flare start | 04:00:00 | 2.5 | 2.0-6.0 (2.0-3.2) | - | 04:00:12 | 4.0 (3.5) |
| | | | start | 04:05:00 | 3.0 | 2.0-7.0 (3.2-5.0) | - | 04:04:20 | 4.0 (3.5) |
| | | | peak | 05:17:00 | 8.8 | 2.0-12.5 (5.0-7.9) | 7.0-10.0 (9.1) | 05:17:28 | 4.0-10.0 (5.1) |
| | | | end | 05:30:00 | 5.0 | 2.0-8.0 (5.0-7.9) | 6.3-9.0 (8.7) | 05:32:52 | 4.0-7.0 (4.6) |
| 25-Dec-2015 | 3 | C2.0 | pre-flare start | 09:15:00 | 6.2 | 2.0-8.0 (5.0-7.9) | - | Saturated XRT images | |
| | | | start | 09:20:00 | 7.65 | 2.0-9.0 (7.9-12.5) | - | | |
| | | | peak | 09:38:00 | 9.76 | 2.0-12.5 (7.9-12.5) | 8.0-10.0 (8.9) | 09:39:58 | 5.0-10.0 (5.6) |
| | | | end | 10:01:00 | 7.0 | 2.0-8.0 (5.0-7.9) | 6.0-8.0 (8.7) | 09:59:58 | 5.0-9.0 (5.2) |
| 25-Dec-2015 | 4 | C1.0 | pre-flare start | 12:28:00 | 4.8 | 2.0-9.0 (3.2-5.0) | - | 12:28:50 | 3.5-5.0 (4.0) |
| | | | start | 12:33:00 | 5.2 | 2.0-9.0 (5.0-7.9) | - | 12:35:20 | 3.5-7.0 (4.2) |
| | | | peak | 12:45:00 | 7.6 | 2.0-10.0 (5.0-7.9) | 7.0-10.0 (9.1) | 12:44:48 | 5.0-10.0 (4.6) |
| | | | end | 13:04:00 | 5.0 | 2.0-6.0 (5.0-7.9) | 7.0-9.0 (8.8) | 13:03:29 | 4.0-5.0 (4.4) |
| 25-Dec-2015 | 5 | C1.1 | pre-flare start | 13:02:00 | 5.6 | 2.0-9.0 (5.0-7.9) | - | 13:03:25 | 4.0-5.0 (4.4) |
| | | | start | 13:07:00 | 5.8 | 2.0-9.0 (3.2-5.0) | - | 13:06:25 | 4.0-6.0 (4.3) |
| | | | peak | 13:14:00 | 7.2 | 2.0-10.0 (5.0-7.9) | 7.0-9.0 (8.8) | 13:15:49 | 5.0-10.0 (4.3) |
| | | | end | 13:25:00 | 4.9 | 2.0-10.0 (5.0-7.9) | 7.0-8.0 (8.8) | 13:22:05 | 3.0-7.0 (4.2) |
| 25-Dec-2015 | 6 | C6.9 | pre-flare start | 16:54:00 | 3.9 | 2.0-5.0 (3.2-5.0) | - | No XRT data data available | |
| | | | start | 16:59:00 | 4.0 | 2.0-8.0 (5.0-7.9) | - | | |
| | | | peak | 17:19:00 | 13.34 | 2.0-12.5 (5.0-7.9) | 7.9-10.0 (9.2) | 17:17:46 | 4.0-10.0 (4.2) |
| | | | end | 17:49:00 | 6.5 | 2.0-9.0 (3.2-5.0) | 7.9-10.0 (9.1) | 17:36:28 | 3.0-4.0 (3.6) |
| 25-Dec-2015 | 7 | C3.0 | pre-flare start | 22:43:00 | 2.6 | 2.0-8.0 (3.2-5.0) | - | 22:41:41 | 3.0-4.0 (3.6) |
| | | | start | 22:48:00 | 4.5 | 2.0-8.0 (3.2-5.0) | - | No XRT data until 23:30 UT | |
| | | | peak | 23:00:00 | 12.1 | 2.0-12.5 (7.9-12.5) | 7.0-10.0 (5-7.9) | No XRT data until 23:30 UT | |
| | | | end | 23:14:00 | 7.0 | 2.0-11.0 (5.0-7.9) | 7.0-8.0 (2-3.2) | 23:30:09 | 4.0-9.0 (4.5) |
| 26-Dec-2015 AR #12473 S22 E09 (βγ) | 8 | C1.6 | pre-flare start | 01:39:00 | 2.5 | 2.0-8.0 (2.0-3.2) | - | 01:38:35 | 3.0-4.0 (3.4) |
| | | | start | 01:44:00 | 3.0 | 2.0-8.0 (3.2-5.0) | - | 01:44:00 | 3.0-5.0 (3.6) |
| | | | peak | 01:53:00 | 10.0 | 2.0-12.5 (7.9-12.5) | 8.0-10.0 (8.9) | 01:57:00 | 6.0-10.0 (5.9) |
| | | | end | 02:32:00 | 5.5 | 2.0-11.0 (5.0-7.9) | 7.0-8.0 (8.5) | No XRT data | |
| 26-Dec-2015 | 9 | C7.8 | pre-flare start | 04:56:00 | 4.0 | 2.0-8.0 (3.2-5.0) | - | 04:57:00 | 3.0-4.0 (3.9) |
| | | | start | 05:01:00 | 3.0 | 2.0-8.0 (3.2-5.0) | - | 05:00:00 | 3.0-4.0 (3.6) |
| | | | peak | 05:10:00 | 14.6 | Saturated AIA data | Saturated AIA data | Saturated XRT data | |
| | | | end | 05:23:00 | 11.0 | Saturated AIA data | Saturated AIA data | No XRT data | |

Table A2: Sigmoid and X-ray flare observation details

| (Col. 1) Date | (Col. 2) Flare no. | (Col. 3) X-ray Class | (Col. 4) Phase of the flare | (Col. 5) Time (UT) | (Col. 6) GOES temp. (MK) | (Col. 7) Temp. from EM (MK) | (Col. 8) AIA filter-ratio temp (MK) | (Col. 9) Time (UT) | (Col. 10) XRT filter-ratio temp (MK) |
|---|---|---|---|---|---|---|---|---|---|
| 26-Dec-2015 | 10 | C1.2 | pre-flare | 11:50:00 | 3.0 | 2.0-8.0 (3.2-5.0) | - | 11:51:40 | 4.0-5.0 (4.1) |
| | | | start | 11:55:00 | 3.5 | 2.0-8.0 (3.2-5.0) | - | 11:54:55 | 4.0-5.0 (4.1) |
| | | | peak | 12:03:00 | 8.3 | 2.0-12.5 (5.0-7.9) | 7.0-10.0 (9.0) | 12:03:44 | 5.0-12.5 (5.4) |
| | | | end | 12:16:00 | 5.5 | 2.0-8.0 (5.0-7.9) | 7.0-10.0 (8.6) | 12:16:00 | 4.0-8.0 (4.6) |
| 26-Dec-2015 | 11 | C4.1 | pre-flare | 15:05:00 | 5.5 | 2.0-8.0 (3.2-5.0) | - | 15:00:34 | 4.0-5.0 (3.9) |
| | | | start | 15:10:00 | 6.0 | 2.0-8.0 (3.2-5.0) | - | 15:09:53 | 4.0-5.0 (4.5) |
| | | | peak | 15:26:00 | 12.5 | Saturated AIA data | Saturated AIA data | | No XRT data |
| | | | end | 15:43:00 | 9.0 | 2.0-8.0 (5.0-7.9) | 8.9 | | No XRT data |
| 28-Dec-2015 AR #12473 S22 W18 ($\beta\delta$) | 12 | M1.8 | pre-flare | 11:15:00 | 4.0 | 2.0-8.0 (3.2-5.0) | - | | No |
| | | | start | 11:20:00 | 4.0 | 2.0-8.0 (3.2-5.0) | - | | XRT |
| | | | peak | 12:08:00 | 14.5 | 2.0-12.5 (7.9-12.5) | 9.0-12.5 (9.5) | | data |
| | | | end | 13:09:00 | 8.8 | 2.0-12.5 (5.0-7.9) | 6.0-12.5 (8.6) | | available |

This paper has been typeset from a TEX/LATEX file prepared by the author.